\documentclass[twocolumn,prl,letterpaper,showpacs]{revtex4}
\usepackage{amsmath}
\usepackage{graphicx}
\usepackage{amssymb}
\usepackage{color}
\usepackage{float}
\usepackage{geometry,amssymb,amsmath,graphicx,textcomp,calc,ifthen,epsfig}
\usepackage{pdfpages}
\usepackage{pgffor}

\begin{document}

\title{Antibunched photons emitted by a dc-biased Josephson junction}

\author{C. \surname{Rolland}$^{1}$}
\thanks{These two authors contributed equally.}
\author{A. \surname{Peugeot}$^{1}$}
\thanks{These two authors contributed equally.}
\author{S. \surname{Dambach}$^{2}$}

\author{M. \surname{Westig}$^{1}$}

\author{B. \surname{Kubala}$^{2}$}

\author{Y. \surname{Mukharsky}$^{1}$}

\author{C. \surname{Altimiras}$^{1}$}

\author{H. \surname{le Sueur}$^{1}$}

\author{P. \surname{Joyez}$^{1}$}

\author{D. \surname{Vion}$^{1}$}

\author{P. \surname{Roche}$^{1}$}

\author{D. \surname{Esteve}$^{1}$}

\author{J. \surname{Ankerhold}$^{2}$}
\email{email: joachim.ankerhold@uni-ulm.de}

\author{F. \surname{Portier}$^{1}$}
\email{email: fabien.portier@cea.fr}

\affiliation{$^{1}$ DSM/IRAMIS/SPEC, CNRS UMR 3680, CEA, Universit\'e Paris-Saclay, 91190 Gif sur Yvette, France}
\affiliation{$^{2}$ Institute for Complex Quantum Systems and IQST, University of Ulm, 89069 Ulm, Germany}

\begin{abstract}
We show experimentally that a dc biased Josephson junction in series with a high-enough-impedance microwave resonator emits antibunched photons.  Our resonator is made of a simple micro-fabricated spiral coil that resonates at 4.4 GHz and reaches a 1.97 k$\Omega$ characteristic impedance. The second order correlation function of the power leaking out of the resonator drops  down to 0.3 at zero delay, which demonstrates the antibunching of the photons emitted by the circuit at a rate of 6 $10^7$ photons per second. Results are found in quantitative agreement with our theoretical predictions. This simple scheme could offer an efficient and bright single-photon source in the microwave domain. 
\end{abstract}

\date{\today}

\maketitle

Single photon sources constitute a fundamental resource for many quantum information technologies, notably secure quantum state transfer using flying photons. In the microwave domain, although photon propagation is more prone to losses and thermal photons present except at extremely low temperature, applications can nevertheless be considered \cite{PhysRevX.7.011035,Nori}. Single microwave photons were first demonstrated in \cite{houck07} using the standard design of single-photon emitters: an anharmonic atom-like quantum system excited from its ground state relaxes by emitting a single photon on a well-defined transition before it can be excited again. The first and second order correlation functions of such a source \cite{Bozyigit2011} demonstrate a rather low photon flux limited by the excitation cycle duration, but an excellent antibunching of the emitted photons. 
We follow a different approach, where the tunnelling of discrete charge carriers through a quantum coherent conductor creates photons in its embedding circuit. The resulting quantum electrodynamics of this type of circuits \cite{PhysRevB.91.205417,PhysRevB.93.075425, PhysRevB.95.125311,PhysRevX.6.031002,PhysRevLett.116.043602,JuhaNJP,PhysRevLett.110.267004} has been shown to provide e.g. masers \cite{PhysRevLett.113.036801,PhysRevLett.114.130403,PhysRevB.90.020506,Cassidy939}, simple sources of non-classical radiation \cite{PhysRevLett.113.043602, PhysRevLett.117.056801,PhysRevLett.119.137001}, or near quantum-limited amplifiers \cite{Jebari18}. When the quantum conductor is a Josephson junction, dc biased at voltage $V$ in series with a linear microwave resonator, exactly one photon is created in the resonator each time a Cooper pair tunnels through the junction, provided that the Josephson frequency $2eV/h$ matches the resonator's frequency \cite{PhysRevLett.106.217005}.We demonstrate here that in the strong coupling regime between the junction and the resonator, the presence of a single photon in the resonator inhibits the further tunneling of Cooper pairs, leading to the antibunching of the photons leaking out of the resonator \cite{PhysRevLett.111.247002,PhysRevB.92.054508}. Complete antibunching is expected when the characteristic impedance of the resonator reaches $Z_c=2 R_Q/\pi$, with $R_Q=h/(2e)^2 \simeq 6.45$\,k$\Omega$ the superconducting resistance quantum. This regime, for which the analogue of the fine structure constant of the problem is of order 1, has recently attracted attention \cite{1802.00633,1805.07379}, as it allows the investigation of many-body physics with photons \cite{PhysRevB.85.140506, PhysRevLett.110.017002} or ultra-strong coupling physics \cite{PhysRevLett.111.243602}, offering new strategies for the generation of non classical radiation \cite{1367-2630-19-2-023036}.

\begin{figure}[!h]
\centering
\includegraphics[width= 7 cm]{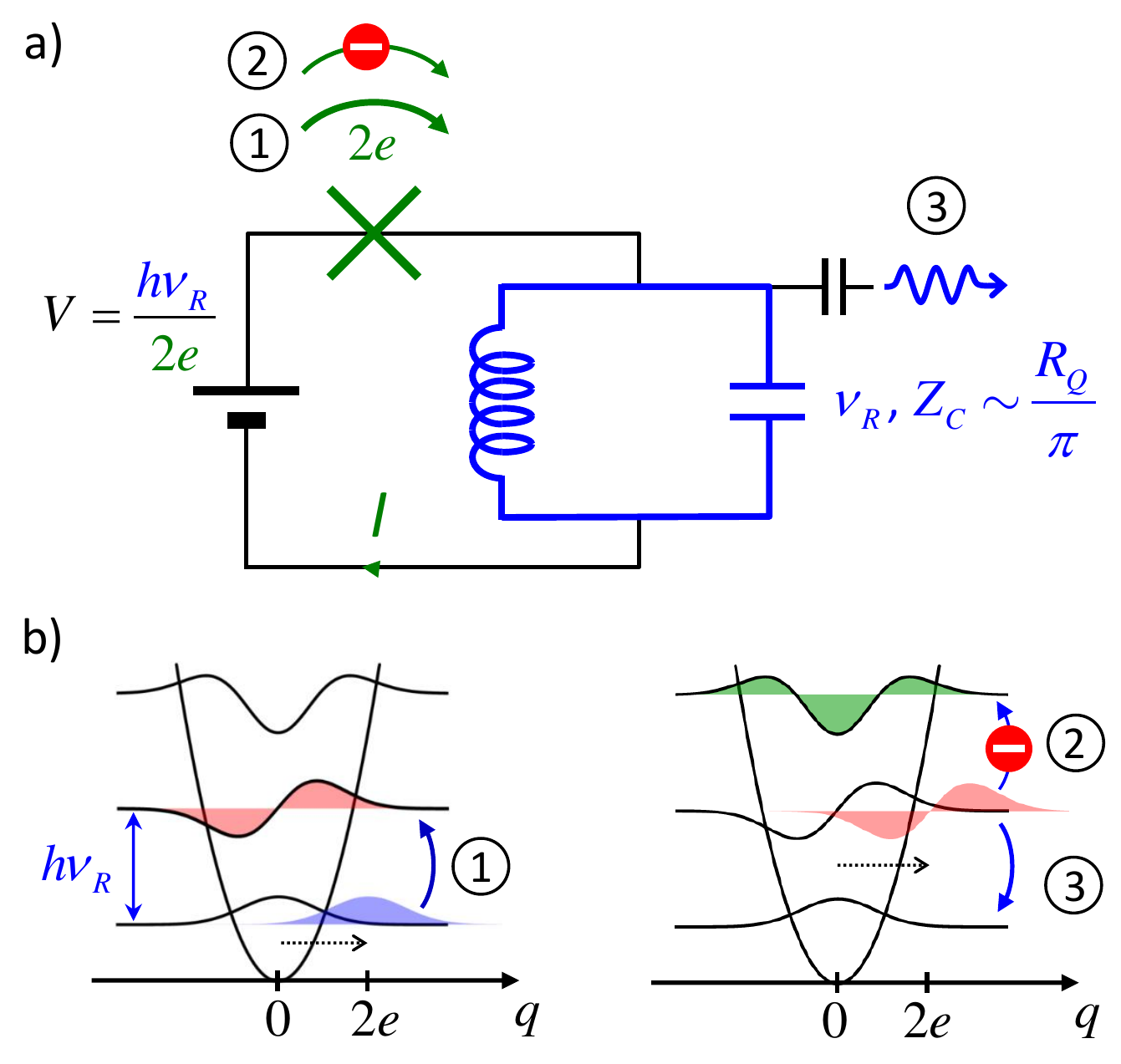}
\caption{\label{fig:principle}
\textbf{Principle of the experiment:}
\textbf{(a)} A Josephson junction in series with a resonator of frequency $\nu_R$ and characteristic impedance $Z_c$ of the order of $R_Q=h/(2e)^2$ is voltage biased so that each Cooper pair that tunnels produces a photon in the resonator (1).  \textbf{(b)} Photon creation and relaxation: A tunneling Cooper pair shifts the charge on the resonator capacitance by $2e$. The tunneling rate $\Gamma_{n\rightarrow n+1}$ starting with the resonator in Fock state $| n \rangle$  is proportional to the overlap between the wavefunction $\Psi_n(q)$ shifted by $2e$ and $\Psi_{n+1}(q)$. This overlap depends itself on $Z_C$ via the curvature of the resonator energy.  At a critical $Z_c$, $\Gamma_{1\rightarrow 2}$ = 0 and no additional photons can be created  (2) until the photon already present has leaked out (3). The photons produced are thus antibunched, which is revealed by measuring the $g^{(2)}$ function of the leaked radiation.}
\end{figure}

The simple circuit used in this work is represented in  Fig.~\ref{fig:principle}a: a Josephson junction is coupled to a microwave resonator of frequency $\nu_R$ and characteristic impedance $Z_c$, and biased at a voltage $V$ smaller than the gap voltage $V_{\mathrm{gap}}=2 \Delta /e$ , where $-e$ is the electron charge and $\Delta$ the superconducting gap, so that single electron tunneling is impossible. The time-dependent Hamiltonian
\begin{equation}
\label{H_tdependent} 
H=(a^\dagger a+1/2)h\nu_R - E_{{\mathrm J}}\cos[\phi(t)]
\end{equation}
of the circuit is the sum of the resonator and Josephson Hamiltonians. Here $a$ is the photon annihilation operator in the resonator, $E_J$ is the Josephson energy of the junction, $\phi(t)=2eVt/\hbar  - \sqrt{r}(a +a^\dagger)$ is the phase difference across the  junction (conjugate to the number of Cooper pairs  transferred accross the junction), and $r= \pi Z_c/R_Q$ is the charge-radiation coupling in this one-mode circuit \cite{ingold92}. The nonlinear Josephson Hamiltonian thus couples Cooper pair transfer to photon creation in the resonator.
This results in inelastic Cooper pair tunneling: a dc current flows in this circuit when the electrostatic energy provided by the voltage source upon the transfer of a Cooper pair corresponds to the energy of an integer number $k$ of photons created in the resonator: $2eV=k h \nu_R$. The steady state occupation number $\bar{n}$  in the resonator results from the balance between the Cooper pair tunneling rate and the leakage rate to the measurement line. For $k= 1$ -- the resonance condition  of the AC Josephson effect -- each Cooper pair transfer creates a single photon. The theory of dynamical Coulomb blockade (DCB)\cite{averin90,ingold92,holst94} predicts that, in the limit of small coupling $r$, the power emitted into an empty resonator 
\begin{equation}
\mathcal{P}=\frac{2 e^2 E_J^{*2}}{\hbar^2} \mathrm{Re} Z(\nu=2eV/h) 
\label{Power}
\end{equation} 
 coincides with the AC Josephson expression, albeit with a reduced effective Josephson energy $E_J^{*}=E_J e^{-r/2}$ renormalized by the zero-point phase fluctuations of the resonator \cite{SM,SCHON1990237, 0295-5075-44-3-360, PhysRevLett.110.217003, PhysRevLett.111.247002,PhysRevB.92.054508,1802.00633}.
In the strong-coupling regime ($r \simeq 1$), however, the single rate description above breaks down as a single photon in the resonator already influences further emission processes, as explained in Fig.~\ref{fig:principle}b.

 A more sophisticated theory \cite{PhysRevLett.111.247002,PhysRevB.92.054508} addressing this regime considers the Hamiltonian \eqref{H_tdependent} in the rotating-wave approximation at the resonance condition $2eV=h\nu_R$ for single photon creation. Expressed in the resonator Fock state basis $\{| n \rangle \}$, $H$ reduces to $H^\textrm{RWA} = -(E_J/2) \sum_n \left( h^\textrm{RWA}_{n,n+1} |n\rangle\langle n+1| + \mathrm{h.\,c.}\right)\,$, with the transition matrix elements
\begin{equation}\label{H_RWA}
h^\textrm{RWA}_{n,n+1}=   \langle n| \exp\left[i\sqrt{r}( a^\dagger +a)\right]  |n+1\rangle.
\end{equation}
Describing radiative losses via a Lindblad super-operator, one gets the second order correlation function for vanishing occupation number $\bar{n}\ll1$  \cite{PhysRevLett.111.247002,PhysRevB.92.054508}: 

\begin{eqnarray}               
g^{(2)}(\tau)=&\displaystyle{\frac{\left\langle a^\dagger(0) a^\dagger(\tau) a(\tau) a(0) \right \rangle}{\left\langle a^\dagger  a \right \rangle^2}}\nonumber\\
=& \left[1- \displaystyle{\frac{r}{2}} \exp{(-\kappa \tau/2)}\right]^2 \label{g2}
\end{eqnarray}
\noindent 
with $\kappa$ the photon leakage rate of the resonator. In the low coupling limit $r\ll1$ where $h^\textrm{RWA}_{n,n+1}$ scales as $\sqrt{n+1}$, one recovers the familiar Poissonian correlations $g^{(2)}(0)=1$. On the contrary, at $r=2$ ($Z_c = 4.1$ k$\Omega$),  $h^\textrm{RWA}_{1,2} = 0$ and Eq. (\ref{g2})  yields perfect antibunching of the emitted photons: $g^{(2)} (0)=0$. In this regime, as illustrated by Fig.~\ref{fig:principle}, a first tunnel event bringing the resonator from Fock state $|0\rangle$ to $|1\rangle$ cannot be followed by a second one as long as the photon has not been emitted in the line. This is the mechanism involved in the Frank-Condon effect and relies on the reduction of the matrix element of the Josephson Hamiltonian between the one and two photon states of the cavity as the coupling parameter $r$ increases from 0 to 2, where it vanishes. It is thus different from the mechanism at work in the recent work of Grimm and coworkers \cite{PhysRevX.9.021016} which relies on the charge relaxation induced by a large on chip resistance.

Standard on-chip microwave resonator designs yield characteristic impedances of the order of $100\thinspace\Omega$, i.e. $r\sim0.05$. To appoach $r\sim1-2$, we have micro-fabricated a resonator with a spiral inductor etched in a 150 nm niobium film sputtered onto a quartz substrate, chosen for its low dielectric constant  ($\epsilon_r \simeq 3.8$), which was then connected to a SQUID loop, of normal resistance $R_t= 222 \pm 3\ \mathrm{k\Omega}$, acting as a flux-tunable Josephson junction (see Fig.~2).
The outgoing radiation was collected in a $50\ \Omega$ line through an impedance-matching stage aiming at lowering the resonator quality factor. The geometry of the resonator was optimized using the microwave solver Sonnet,  predicting a resonant frequency $\nu_R = 5.1 \thinspace $GHz, with a characteristic impedance of 2.05\ k$\Omega$, corresponding to $r=1.0$, and a quality factor $Q=2 \pi \nu_r/\kappa = 42$  \cite{SM}. 
The actual values measured using the calibration detailed in the Supplemental material \cite{SM} are $\nu_r$ = 4.4 GHz, $Q= 36.6$, and a characteristic impedance $Z_c = 1.97 \pm 0.06\ \mathrm{k\Omega}$, corresponding to a coupling parameter $r=0.96 \pm 0.03$, and thus to an expected $E_J^*/E_J=0.62\pm0.01$. We attribute the small difference between design and experimental values to a possible under-estimation in our microwave simulations of the capacitive coupling of the resonator to the surrounding grounding box.

The sample is placed in a shielded sample holder thermally anchored to the mixing chamber of a dilution refrigerator at $T=$12 mK. As shown in Fig.~2, the sample is connected to a bias tee, with a dc port connected to a filtered voltage divider, and an rf port connected to a 90$^o$ hybrid coupler acting as a microwave beam splitter towards two amplified lines with an effective noise temperature of 13.8 K. After bandpass filtering at room temperature, the signals in these two channels $V_a(t),V_b(t)$ are down converted to the 0 - 625 MHz frequency range using two mixers sharing the same local oscillator at $\nu_{\mathrm{LO}}= 4.71$ GHz, above the  resonator frequency. The ouput signals are then digitized at  1.25~GSamples/s to measure their two quadratures, and the relevant correlation functions are computed numerically.

\begin{figure}
\centering
\includegraphics[width= 7 cm]{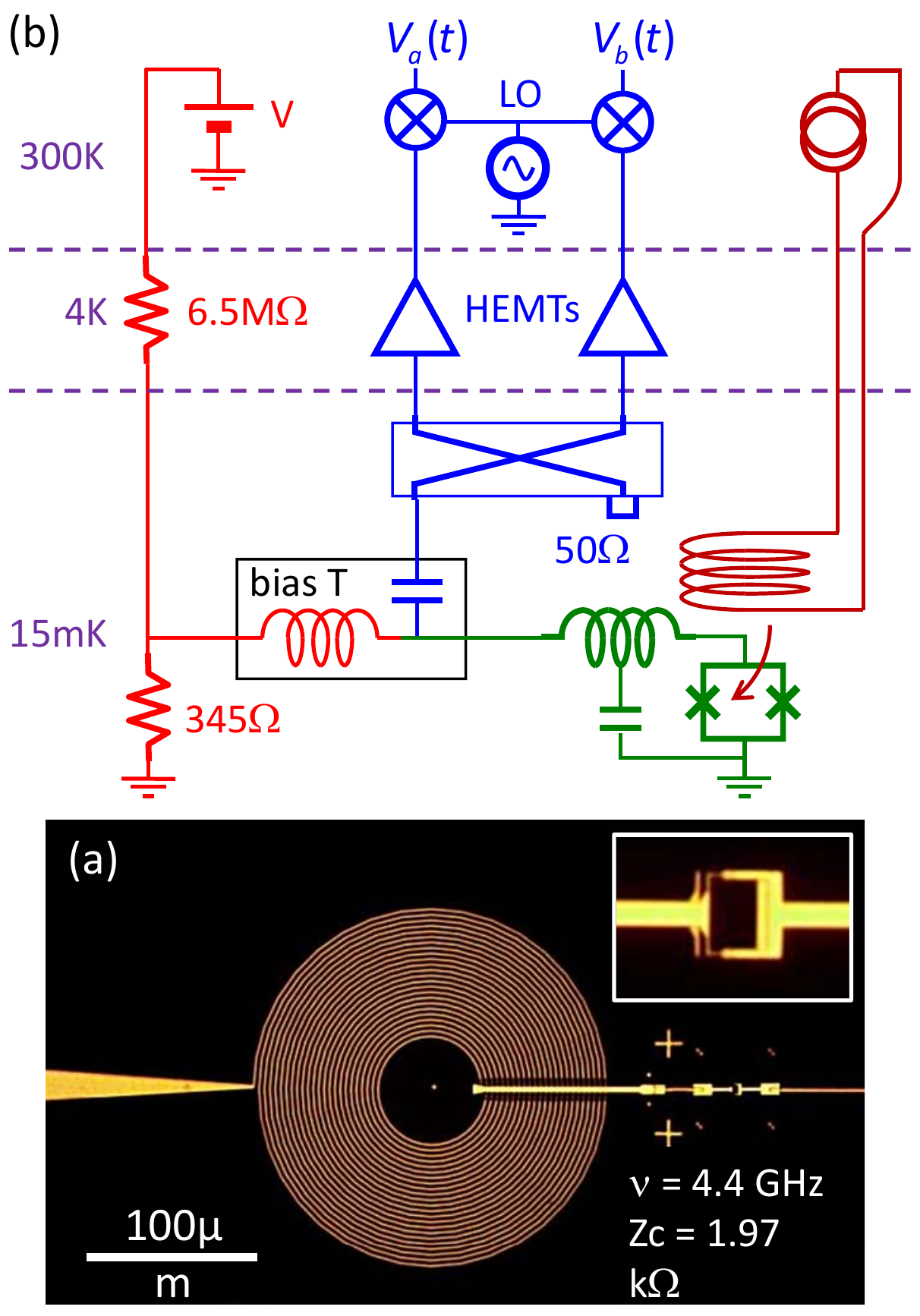}
\caption{\label{fig:scheme}
\textbf{Experimental setup. (a)} Optical micrography of the sample showing the Al/AlOx/Al SQUID (inset) implementing the Josephson junction and the resonator made of a Nb spiral inductor with stray capacitance to ground. \textbf{(b)} Schematic of the circuit showing the sample (green), the coil circuit for tuning the Josephson energy (brown),  the dc bias line (red), and the bias tee connected to the microwave line (blue) with bandpass filters, isolators (not shown here), and a symmetric splitter connected to two measurement lines with amplifiers at 4.2\ K and demodulators at room temperature \cite{SM}. 
}
\end{figure}

\begin{figure}[!b]
\centering
\includegraphics[width= 7 cm]{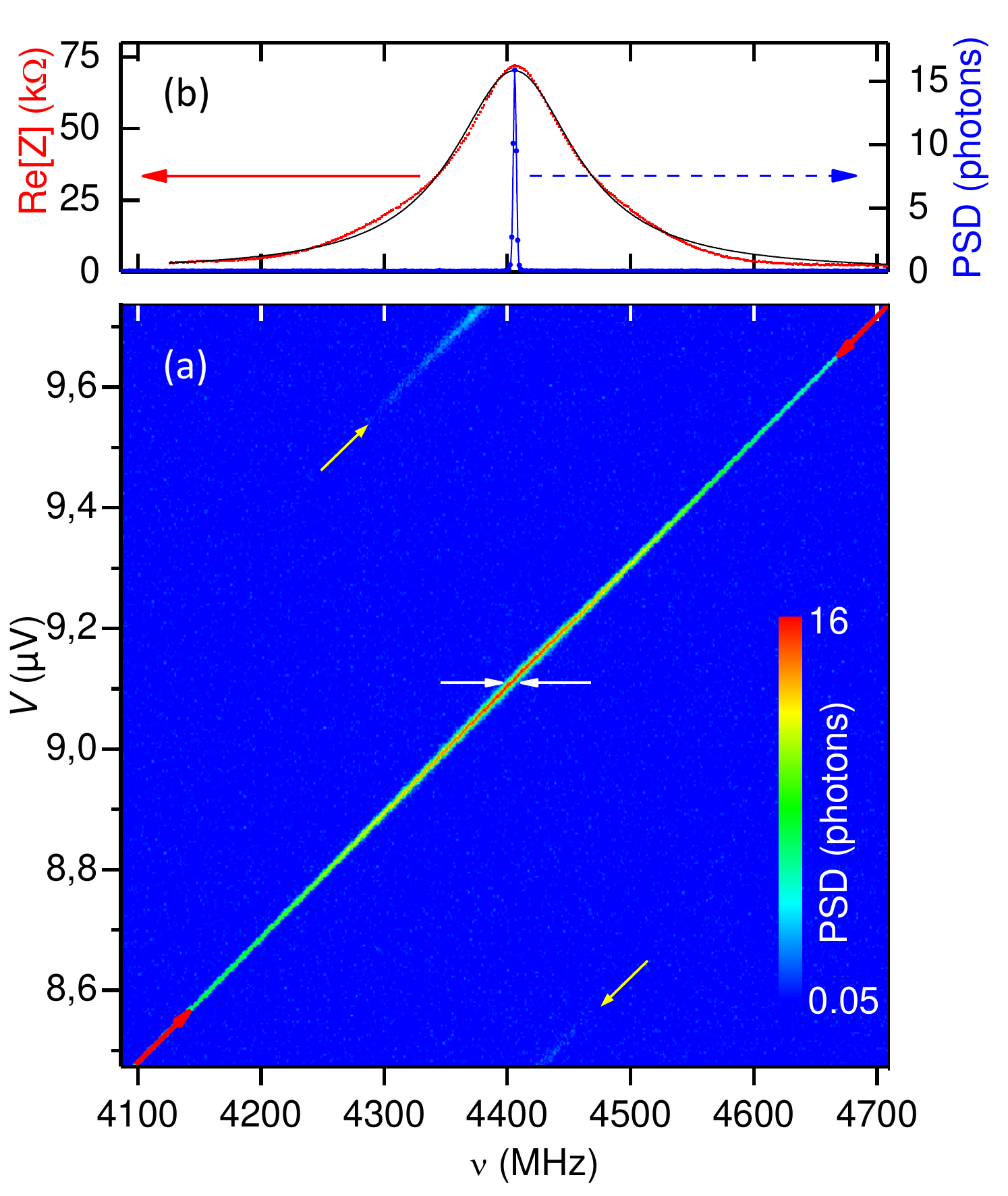}
\caption{\label{fig:2DMap}
\textbf{Emitted microwave power and impedance seen by the junction. (a)} 2D map of the emitted power spectral density (PSD) as a function of the frequency $\nu$ and bias voltage $V$, expressed in photon occupation number (logarithmic color-scale). \textbf{(b)} Spectral line at $V = 9.11\,\mu V$ (blue points) obtained from a cut in the 2D map along the horizontal white arrows and  real part of the impedance $\mathrm{Re}[Z(\nu)]$ seen by the SQUID (red points). The solid blue (black) line is a Gaussian (Lorentzian) fit.}
\end{figure}

In Fig.~\ref{fig:2DMap}a, the  measured 2D emission map as a function of bias voltage and frequency shows  the single photon regime along the diagonal. A cut at the resonator frequency (blue line in Fig.~3b) reveals an emission width of 2.9 MHz, which we attribute to low frequency fluctuations of the bias voltage, mostly of thermal origin. Two faint lines (pointed by the oblique yellow arrows) also appear at $2eV = h (\nu \pm \nu_P)$, and correspond to the simultaneous emission of a photon in the resonator and the emission/absorption of a photon in a parasitic resonance of the detection line at $\nu_P=$ 325\,MHz. Comparing the weight of these peaks to the main peak at $2eV=h\nu$ yields a $61\ \Omega$ characteristic impedance of the parasitic mode and a $15$ mK mode temperature in good agreement with the refrigerator temperature. 

\begin{figure*}[t]
\centering
\includegraphics[width= 15 cm]{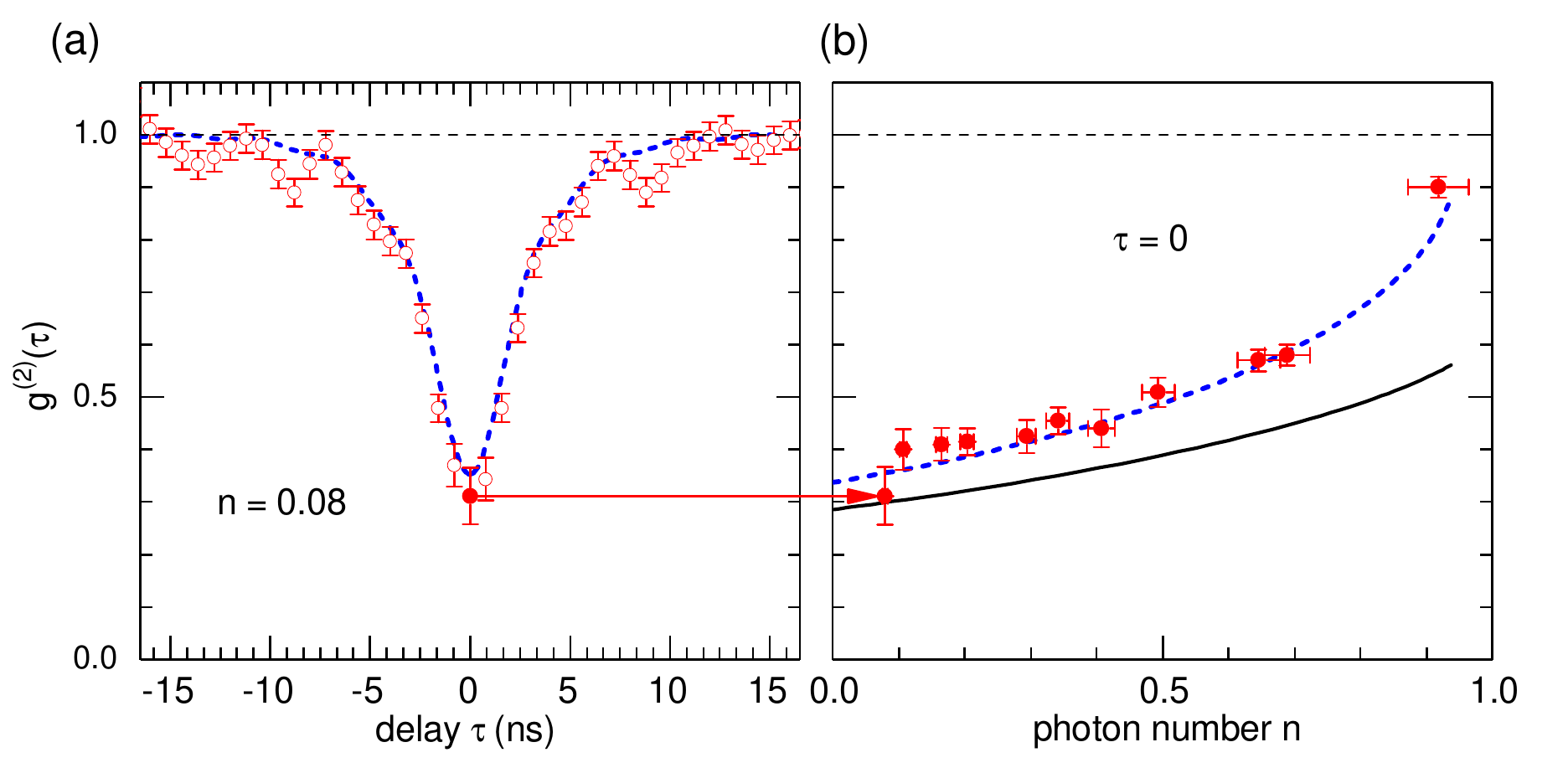}
\caption{\label{fig:g2}
\textbf{Antibunching of the emitted radiation} at bias $V = h \nu_R/2e= 9.11\,\mu$V. \textbf{(a)} Experimental (dots) and theoretical (dashed line) second order correlation function $g^{(2)}$ as a function of delay $\tau$ for $n=0.08$ photons in the resonator. Error bars indicate $\pm$ the statistical standard deviation.
\textbf{(b)} Experimental (dots) and theoretical (dashed line) $g^{(2)}(0)$ as a function of $n$. The solid line is the theoretical prediction not taking into account the finite detection bandwidth.}
\end{figure*}

We now set the bias at $V=h \nu_r/2e=9.1\ \mu$V, and we detect the output signals of the two amplifiers in a frequency band of 525 MHz ($\sim$ 4.4 resonator's FWHM) centered at the emission frequency $\nu_R$. This apparently large detection window -- 180 times wider than the emission line, see Fig.~3b -- is actually barely  enough to measure the fast fluctuations occuring at frequencies up to the inverse resonator lifetime. An even larger bandwidth would   bring  the measured $g^{(2)}$ closer to the expected value of Eq. (\ref{g2}) but would also increase the parasitic fluctuations due to amplifiers' noise and increase the necessary averaging time. Our choice is thus a compromise,  leading to a 15-day long averaging for the lowest occupation number. From the down-converted signals, we rebuild their complex envelopes $S_{a,b}(t)$ \cite{PhysRevA.82.043804, SM}. 
We now use two alternative methods to extract $g^{(2)}(\tau)$. First, we obtain the instantaneous powers $P_{a,b}(t)=|S_{a,b}(t)|^2$, and extract
\begin{equation}
g^{(2)}(\tau)=\frac{\left\langle P_a(t) P_b(t+\tau)\right\rangle}{\left\langle P_a (t)\right\rangle \left\langle P_b(t+\tau)\right\rangle}  
\end{equation}

from their cross-correlations.
Here, the sample's weak contribution has to be extracted from the large background noise of the amplifiers, which we measure by setting the bias voltage to zero. To overcome this complication and get a better precision on $g^{(2)}$, we compute the complex cross-signal $C(t)={S_a}^*(t)S_b(t)$, which is proportional to the power emitted by the resonator and has a negligible background average contribution. $g^{(2)}(\tau)$ can then be extracted from the correlation function of $C(t)$ and $C^*(t)$ \cite{SM}. As $g^{(2)}(\tau)$ is real and the instantaneous noise on $C(t)$ is spread evenly between real and imaginary parts, this also improves the signal to noise ratio by $\sqrt{2}$.

Both methods gave the same results within their standard deviations, and the $g^{(2)}$ values shown in Fig.~4 correspond to the average of the two procedures.  As we decrease the photon emission rate by adjusting $E_J$ with the magnetic flux threading the SQUID, $g^{(2)}(0)$ decreases. For the lowest measured emission rate of $60$ millions photons per second, corresponding to an average resonator population of 0.08 photons, $g^{(2)}(0)$ goes down to 0.31$\pm 0.04$, in good agreement with the theoretical prediction of 0.27, cf. Eq. (4) for r=0.96. This is the main result of this work, which demonstrates a significant antibunching of the emitted photons. In agreement with Eq. (\ref{g2}), the characteristic time scale of the $g_2(\tau)$ variations coincides with the 1.33 ns resonator lifetime deduced from the calibrations. As our design did not reach $r=2$, the transition from $|1\rangle$ to $|2\rangle$ is not completely forbidden, and from then on, transitions from $|2\rangle$ to $|3\rangle$ and higher Fock states can occur. The larger $E_J$, the more likely to have 2 photons and hence photon bunching. To predict the time-dependent $g^{(2)}(\tau)$ for arbitrary $E_J$, we solve the full quantum master equation
\begin{equation}
\dot{\rho} = -\frac{i}{\hbar}[H^\textrm{RWA},\rho]+ \frac{\kappa}{2} \left(2 a \rho a^\dagger - a^\dagger a \rho-\rho a^\dagger a\right)\textcolor{red}{.}
\label{Lindblad}
\end{equation}
This approach also allows for the quantitative modeling of the experimental measurement via a four-time correlator \cite{SM}. Properly accounting for filtering in the measurement chain (see Ref.~\cite{PhysRevA.82.043804,Bozyigit2011} and Supplemental Material \cite{SM}), this description accurately reproduces the experimental results in Fig.~\ref{fig:g2} (lines) without any fitting parameters.

We finally probe the renormalization of $E_J$ by the zero point fluctuations of the resonator using Eq. (\ref{Power}). This requires to maintain the resonator photon population much below 1, which should be obtained by reducing the Josephson energy using the flux through the SQUID. However, magnetic hysteresis due to vortex pinning in the nearby superconducting electrodes prevented us from ascribing a precise flux to a given applied magnetic field, the only straightforward and reliable working point at our disposal thus occurring at zero magnetic flux and maximum Josephson energy. To ensure that the SQUID remains in the DCB regime even at this maximum  $E_J$,  and ensure a low enough photon population, we select a bias voltage $V=10.15\ \mu$V  yielding  radiation at 4.91 GHz, far off the resonator frequency. Here again, the normal current shot noise is used as a calibrated noise source to measure in-situ $G \mathrm{Re}Z(\nu=4.91 \ \mathrm{GHz})$. The effective Josephson energy $E_J^*=1.86\pm0.02\ \mu$eV extracted in this way is significantly smaller than the Ambegaokar-Baratoff value of $E_J=3.1\pm0.03\ \mu$eV, and in good agreement with our prediction of  $E_J^*=1.84\pm0.03\ \mu$eV \cite{noteEJ}, taking also into account the phase fluctuations coming from the parasitic mode at $\nu_p$ and its harmonics.

In conclusion, we have explored a new regime of the quantum electrodynamics of coherent conductors by strongly coupling a dc biased Josephson junction to its electromagnetic environment, a high-impedance microwave resonator. This enhanced coupling first results in a sizeable renormalization of the effective Josephson energy of the junction. Second, it provides an extremely simple and bright source of antibunched photons. Appropriate time shaping either of the bias voltage \cite{PhysRevA.93.060301}, or the resonator frequency, or the Josephson energy \cite{PhysRevX.9.021016} should allow for on-demand single photon emission. This new regime that couples quantum electrical transport to quantum electromagnetic radiation opens the way to new  devices for quantum microwaves generation. It also allows many fundamental experiments like investigating high photon number processes, parametric transitions in the strong coupling regime \cite{PhysRevB.86.054514,PhysRevB.92.174532,PhysRevLett.111.247002, PhysRevB.92.054508},  the stabilization of a Fock state by dissipation engineering \cite{PhysRevA.93.060301}, or the development of new type of Qbit based on the Lamb-shift induced by the junction \cite{Esteve18}. 

\begin{acknowledgments} 
We thank B. Huard, S. Seidelin and M. Hofheinz for useful discussions. This work received funding from the European Research Council under the European Union’s Programme for Research and Innovation (Horizon 2020)/ERC Grant Agreement No. [639039]. We gratefully acknowledge partial support from LabEx PALM (ANR-10-LABX-0039-PALM), ANR contracts ANPhoTeQ and GEARED, from the ANR-DFG Grant JosephCharli, and from the ERC through the NSECPROBE grant, from IQST and the German Science Foundation (DFG) through AN336/11-1. S.D. acknowledges financial support from the Carl-Zeiss-Stiftung.    
\end{acknowledgments}



\section*{Supplementary material}

# Antibunched photons emitted by a dc-biased Josephson junction: Supplementary Material

C. Rolland[1],* A. Peugeot[1],* M. Westig[1], B. Kubala[2], S. Dambach[2], Y. Mukharsky[1], C. Altimiras[1], H. Lesueur[1], P. Joyez[1], D. Vion[1], P. Roche[1], D. Esteve[1], J. Ankerhold[2],† and F. Portier[1]‡

[1] *SPEC (UMR 3680 CEA-CNRS), CEA Paris-Saclay, 91191 Gif-sur-Yvette, France and*
[2] *Institute for Complex Quantum Systems and IQST,*
*University of Ulm, 89069 Ulm, Germany*

## I. DERIVATION OF EQ. 2 OF THE MAIN TEXT USING $P(E)$ THEORY

The spectral density of the emitted radiation is given by [1]:

$$\gamma(V, \nu) = \frac{2\text{Re}[Z(\nu)]}{R_Q} \frac{\pi}{2\hbar} E_J^2 P(2eV - h\nu), \tag{1}$$

where $Z(\nu)$ is the impedance across the junction, $R_Q$ is the superconducting resistance quantum $R_Q = h/4e^2$, $E_J$ is the Josephson energy of the junction, and $P(E)$ represents the probability density for a Cooper pair tunneling across the junction to dissipate the energy $E$ into the electromagnetic environment described by $Z(\nu)$ [2]. $P(E)$ is a highly nonlinear transform of $Z(\nu)$:

$$\begin{aligned}
P(E) &= \frac{1}{2\pi\hbar} \int_{-\infty}^{\infty} \exp[J(t) + iEt/\hbar] dt \\
J(t) &= \int_{-\infty}^{+\infty} \frac{d\omega}{\omega} \frac{2\text{Re}Z(\omega)}{R_Q} \frac{e^{-i\omega t} - 1}{1 - e^{-\beta\hbar\omega}},
\end{aligned} \tag{2}$$

where $\beta = 1/k_B T$. For an $LC$ oscillator of infinite quality factor at zero temperature, $P(E)$ is given by

$$P(E) = e^{-r} \sum_n \frac{r^n}{n!} \delta(eV - n\hbar\omega_0) \tag{3}$$

where $r = \pi\sqrt{L/C}/R_Q$ and $\omega_0 = 1/\sqrt{LC}$.

Here, we consider the case of a mode of finite linewidth, so that near the resonance the real part of the impedance can be approximated as

$$\frac{2\text{Re}Z(\omega)}{R_Q} \simeq r\mathcal{L}(\omega, \omega_0, Q). \tag{4}$$

where

$$\mathcal{L}(\omega, \omega_0, Q) \equiv \frac{2}{\pi} \frac{Q}{1 + 4Q^2 \left(\frac{\omega}{\omega_0} - 1\right)^2}$$

————

*These two authors contributed equally.
†Electronic address: email:joachim.ankerhold@uni-ulm.de
‡Electronic address: email:fabien.portier@cea.fr

denotes a Lorentzian function centered at $\omega_0$ with a maximum value $\frac{2}{\pi}Q$ and a quality factor $Q = \frac{\omega_0}{\Delta\omega}$. Note that $\int \mathcal{L}(\omega, \omega_0, Q)d\omega = \omega_0$.

For such a finite-$Q$ mode, we aim to get a formula similar to Eq. 3, i.e. we look for an expansion

$$P(E) = P_0(E) + P_1(E) + P_2(E) + \ldots + P_n(E) + \ldots \tag{5}$$

where each $P_n(E) \propto r^n$. However, from the integral expressions (2), accessing the different multiphoton peaks, i.e. calculating $P(E \simeq n\hbar\omega_0)$ is not straight-forward. Such an expansion can be obtained using the so-called Minnhagen equation [2], which is an exact integral relation obeyed by $P(E)$, valid for any impedance. We first establish the Minnhagen equation starting from

$$e^{J(t)} - e^{J(\infty)} = \int_{-\infty}^{t} d\tau J'(\tau)e^{J(\tau)} \; ,$$

which, using the definition (2) of $J$ can be recast as

$$e^{J(t)} - e^{J(\infty)} = -i\int_{-\infty}^{+\infty} d\omega' h(\omega') \int_{-\infty}^{\infty} d\tau e^{-i\omega'\tau} e^{J(\tau)} \theta(t-\tau)$$

where $\theta$ is the Heaviside function, $h(\omega) = \frac{1}{1-e^{-\beta\hbar\omega}} \frac{2\mathrm{Re}Z(\omega)}{R_Q}$ and using the fact that $J(-\infty) = J(\infty)$. The rightmost integral being the Fourier transform of a product, we replace it by the convolution product of the Fourier transforms and use the detailed balance property of $h$ and $P$ to simplify the r.h.s.:

$$e^{J(t)} - e^{J(\infty)} = -i\int_{-\infty}^{+\infty} d\omega' h(\omega') \int du \left( \pi\delta(u) + \frac{ie^{it'u}}{u} \right) P(-\omega' - u)$$

$$= \int_{-\infty}^{+\infty} d\omega' h(\omega') \int du \frac{e^{itu}}{u} P(-\omega' - u).$$

Finally, we take the Fourier transform on both sides and rearrange, which yields the Minnhagen equation

$$P(E) = \frac{\hbar}{E} \int P(E - \hbar\omega) \frac{1}{1-e^{-\beta\hbar\omega}} \frac{2\mathrm{Re}Z(\omega)}{R_Q} d\omega + \delta(E)e^{\mathrm{Re}\,J(\infty)} \; . \tag{6}$$

At zero temperature $\frac{1}{1-e^{-\beta\hbar\omega}} \to \theta(\omega)$ and $P(E)$ is zero for negative energies, so that the Minnhagen equation is most frequently found written as

$$P(E) = \frac{\hbar}{E} \int_0^E P(E - \hbar\omega) \frac{2\mathrm{Re}Z(\omega)}{R_Q} d\omega + \delta(E)e^{\mathrm{Re}\,J(\infty)} \; . \tag{7}$$

Plugging the expansion (5) into Eq. 6, one immediately gets

$$P_0(E) = \delta(E)e^{J(\infty)}$$

$$P_1(E) = \frac{1}{E} \int_{-\infty}^{\infty} P_0(E - \hbar\omega) \frac{r\mathcal{L}(\omega, \omega_0, Q)}{1-e^{-\beta\hbar\omega}} d\hbar\omega$$

$$\simeq \frac{e^{J(\infty)}}{\hbar\omega_0} r\mathcal{L}\left(\frac{E}{\hbar}, \omega_0, Q\right)$$

where the approximation of the last line was obtained assuming that $k_B T \ll \hbar\omega_0$ and taking the value of the denominator at $E = \hbar\omega_0$ –where $\mathcal{L}$ (and $P_1$) peak– which is reasonable if the $Q$ is large

enough. By repeated replacement in Eq. 6 and with similar approximations, one systematically obtains the higher orders terms of (5) as shifted Lorentzians of constant $Q$

$$P_{n \geqslant 1}(E) \simeq e^{J(\infty)} \frac{r^n}{nn!} \frac{\mathcal{L}(E/\hbar, n\omega_0, Q)}{\hbar\omega_0}$$

whose value at each peak are

$$P_{n \geqslant 1}(E = n\hbar\omega_0) = \frac{2}{\pi} e^{J(\infty)} \frac{r^n}{nn!} \frac{Q}{\hbar\omega_0}$$

yielding a tunneling rate at the peaks

$$\Gamma_{2e}(eV = n\hbar\omega_0) = \frac{1}{\hbar} \frac{E_J^2 e^{J(\infty)}}{\hbar\omega_0} \frac{r^n}{n!} \frac{Q}{n}.$$

Note that the Cooper pair rates at different orders scale with an extra $Q/n$ compared to the naive rates obtained from Eq. 3.

In the main text, $E_J^2 e^{J(\infty)}$ is called $E_J^{*2}$. This renormalization of the Josephson energy is obtained from the zero point phase correlator

$$J(\infty) = -\langle \varphi(0)\varphi(0) \rangle = -\int_0^{+\infty} \frac{d\omega}{\omega} \frac{2\mathrm{Re}Z(\omega)}{R_Q} \coth\frac{\beta\omega}{2}$$

which in the limit of $k_B T = 0$ and for an RLC parallel resonator (it is important that $\mathrm{Re}Z(\omega \sim 0) \propto \omega^2$ for proper convergence) yields

$$J(\infty) = -\frac{Qr\left(1 + \frac{2}{\pi}\mathrm{atan}\frac{2Q^2-1}{\sqrt{4Q^2-1}}\right)}{\sqrt{4Q^2-1}} = -r\left(1 - \frac{1}{\pi Q} + \mathcal{O}\left(\frac{1}{Q^2}\right)\right),$$

in agreement with the expression $E_J^* = E_J e^{-r/2}$ used in the main text (The finite-$Q$ correction to this renormalization is of order of 1%, beyond the precision of our measurements). In ref. [1], $E_J^{*2}$ was given with an approximate first-order expansion of the phase correlator valid for small phase fluctuations (and which was correct for the small $r$ value in that paper).

We can use the above expressions to calculate the total emitted power via the single photon processes by two different ways. First, we use Eq. 1 at lowest order, to get the spectral density of the emitted radiation:

$$\gamma(V, \nu) \simeq \frac{2\mathrm{Re}[Z(\nu)]}{R_Q} \frac{\pi}{2\hbar} E_J^2 P_0(E = 2eV - h\nu) = e^{J(\infty)} \frac{2\mathrm{Re}[Z(\nu)]}{R_Q} \frac{\pi}{2\hbar} E_J^2 \delta(2eV - h\nu), \quad (8)$$

which, upon integrating over $\nu$, gives Eq. 2 of the main text. Alternatively, one can calculate the Cooper pair tunneling rate using $P_1$, and get the photon emission rate from energy conservation, yielding the same result.

In Figure I we compare the exact $P(E)$ result and the approximate formula, for the experimental parameters.

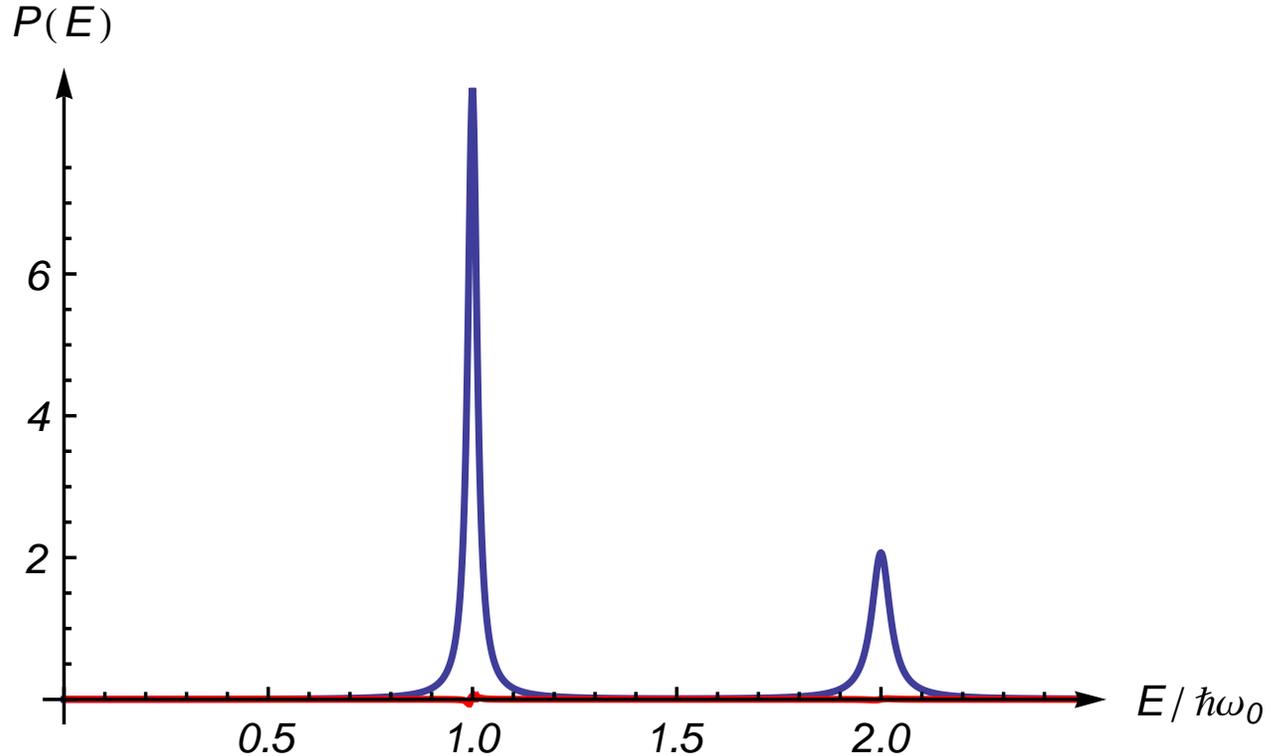

Figure 1: Comparison the exact $P(E)$ result obtained by numerical evaluation of Eqs. (2) and the approximate sum of Lorentzians, evaluated for the experimental parameters ($Q = 36.6$, $r = 0.96$). At this scale, the two curves are indistinguishable. The red curve is the difference between the approximate and the exact result.

## II.   FRANCK-CONDON BLOCKADE IN THE JOSEPHSON-PHOTONICS HAMILTONIAN

The starting point of our theoretical description, the time-dependent Hamiltonian, see Eq. 1 of the main text,

$$H = (a^\dagger a + 1/2)h\nu_R - E_J(\phi)\cos[2eVt/\hbar - \sqrt{r}(a + a^\dagger)],$$  (9)

describes a *harmonic* oscillator with an unusual, *nonlinear* drive term. Going into a frame rotating with the driving frequency, $\omega_J = 2eV/\hbar$, the oscillator operators, $a$ and $a^\dagger$, acquire phase terms rotating with the same frequency. The cosine term of the Hamiltonian can then be rewritten in Jacobi-Anger form so that Bessel functions of order $k$ appear as prefactors of terms rotating with integer multiples of the driving frequency, $k\omega_J$.

A rotating-wave approximation neglects time-dependent terms and, taking proper account of the

commutation relations of oscillator operators, results in the RWA Hamiltonian (on resonance),

$$H^{\mathrm{RWA}} = iE_{\mathrm{J}}e^{-r/2} \; : \; (a^\dagger - a)\frac{J_1(\sqrt{4rn})}{\sqrt{n}} \; : \, ,$$
(10)

where : ... : prescribes normal ordering. While the appearance of a Bessel function highlights the nonlinear-dynamical aspects of the system, the Hamiltonian (10) is completely equivalent to expression Eq. 2 of the main text, given in the main text, using the displacement operator, which emphasizes the connection to Franck-Condon physics.

From either of the two equivalent forms of the RWA Hamiltonian, explicit expressions for the transition matrix elements in terms of associated Laguerre polynomials,

$$h^{\mathrm{RWA}}_{n,n+1} = \frac{ie^{-r/2}\sqrt{r}}{\sqrt{n+1}} \, L^1_n(r) \, ,$$
(11)

can easily be found. Normal ordering reduces the power series of the Bessel function to a low-order polynomial in $r$ (with order $n$ for $h^{\mathrm{RWA}}_{n,n+1}$) and a universal prefactor, describing renormalization of the Josephson coupling. Transition matrix elements thus vanish at the roots of the associated Laguerre polynomials (which in the semiclassical limit of small $r$ and large $n$ approach zeros of the Bessel function $J_1$).

Some simple results can be directly read off from the transition matrix elements; such as the zero-delay correlations that for weak driving measure the probability of two excitations, $g^{(2)}(0) = \langle n(n-1)\rangle/\langle n\rangle^2 \approx 2P_2/P_1^2 \approx \frac{1}{2}\left|h^{\mathrm{RWA}}_{1,2}/h^{\mathrm{RWA}}_{0,1}\right|^2$. The last approximate equality expresses the probabilities $P_{1/2}$ by transition matrix elements, as found by considering the transition rates for the corresponding two-stage excitation process and decay from the Fock states. As mentioned in the main text, in the harmonic limit, $r \ll 1$, where the matrix elements scale with $\sqrt{n+1}$, this would result in the familiar Poissonian correlations and $g^{(2),\mathrm{H0}}(0) = 1$. This contrasts to the antibunching found in our experiment relying on the fact that the experimental parameter $r \sim 1$, while not quite close to the zero of the transition matrix element $h^{\mathrm{RWA}}_{1,2} \propto L^1_1(r) = 2 - r$, is sufficiently large for a considerable suppression of excitations beyond a single photon in the resonator. Coincidentally, the actual value of $r$ is very close to one of the roots, $r \approx 0.93$ of $L^1_3(r) = \frac{1}{6}(-r^3 + 12r^2 - 36r + 24) \propto h^{\mathrm{RWA}}_{3,4}$, so that the system closely resembles a four-level system. In the idealized model description by the approximated Hamiltonian Eq. 3 of the main text and the quantum master equation Eq. 6 of the main text, the vanishing of a transition matrix element implies a strict cut-off of the system's state space at the corresponding excitation level. Various correction terms discussed in the next subsection can lift such a complete blockade and therefore gain relevance once the system is closer to a root than for our $r \sim 1$ value.

## III. CORRECTION TERMS TO HAMILTONIAN AND QUANTUM MASTER EQUATION

The possible impact of various terms and processes not included in RWA Hamiltonian Eq. 3 and quantum master equation Eq. 6 of the main text were carefully checked and found to be completely negligible compared to the error bars due to other experimental uncertainties.

Specifically, the impact of rotating-wave corrections to the time-independent RWA Hamiltonian is sufficiently reduced by the quality factor, $Q = 36.6$. Close to the complete suppression of resonant $g^{(2)}(0)$ contributions at $r = 2$, for very weak driving, and for a bad cavity such processes can become

more relevant, as discussed in some detail in Ref. [3]. The limit of extremely strong driving, where $E_J \gtrsim h\nu_R$, not reached here, is discussed in Appendix D of Ref. [4].

Access to higher Fock-states cut-off by vanishing transition matrix elements could, in principle, also be provided by thermal excitations, i.e., by Lindblad terms not included in the $T = 0$ limit of the quantum master equation Eq. 6 of the main text. The latter, however, is safe to use for the experimentally determined mode temperature of $\sim 15$ mK in our device.

Finally, there are low-frequency fluctuations of the bias voltage, causing the spectral broadening of the emitted radiation (as argued in the main text) that are not accounted for by the quantum master equation Eq. 6 of the main text. Their effect can be modeled, either by an additional Lindblad-dissipator term acting on a density matrix in an extended JJ-resonator space (as described, for instance, in the supplementals to [5]), or by employing the quantum master equation Eq. 6 of the main text and average the results over a (Gaussian) bias-voltage distribution centered around the nominal biasing on resonance.

As argued above, the principal antibunching effect can be understood from transition rate arguments so that it is not sensitive to the phase of the driving, which is becoming undetermined due to the fluctuating voltage bias. In consequence, the measured $g^{(2)}(0)$ is nearly insensitive to low-frequency fluctuations. Residual effects of detuning on $g^{(2)}(\tau)$ entering $g^{(2)}(0)$ via the filtering are negligible due to the large ratio between inverse resonator lifetime and spectral width, cf. Fig. 3 of the main text.

## IV. ACCOUNTING FOR FILTERING

A theoretical approach based on the quantum master equation Eq. 6 of the main text gives direct access to any properly time-/anti-time-ordered products of multiple system operators, which are evaluated by the quantum-regression method. Using input-output theory [6], any arbitrarily ordered product of multiple output operators can readily be expressed in such system-operator objects.

The measured signals, however, do not immediately correspond to output operators but contain operators at the end of the microwave output chain, hence, undergoing additional filtering. An end-of-chain operator acting at a certain time is consequently linked to output operators at all preceding times via a convolution with the filter-response function in the time domain, see the discussion in [7]. Specifically, the measured two-time correlator $G^{(2)}(t_1, t_2) = \langle a_{t_1}^\dagger a_{t_2}^\dagger a_{t_2} a_{t_1} \rangle$, where two operators each are acting at two different times $t_{1/2}$, is related to a four-operator object with each operator acting at a different time.

To simulate the measured $G^{(2)}(t_1, t_2)$, it is necessary to calculate corresponding four-operator objects and then average each instance of time with a probability distribution given by the filter-response function in the time domain. An explicit, worked out example for a three-time object can be found in Appendix E of Ref. [8]. For the special case of a Lorentzian filter-response function a simpler scheme has been put forward [9–11].

For numerical efficiency, here, we calculate the various four-time objects by evaluating the time evolution governed by the exponential of the Liouville superoperator using Sylvester's formula and Frobenius covariants. This approach is completely equivalent to time-evolving the quantum master equation with any standard differential-equation solver. In a final step, three-dimensional temporal integrals have to be numerically evaluated, wherein the uncertainty of time differences (between points at which the different operators act) is linked to the experimental filter function, see above.

The multiple integrals over differently ordered operator objects hamper an intuitive understanding of the effects of filtering. Therefore, it may be helpful to compare the complex effects of filtering

here to a more conventionally encountered scheme describing detection-time uncertainties. In Fig. 2, we show the results of a simple, incomplete filtering description, which only allows for variations in the time difference, $\tau = t_2 - t_1$, but artificially keeps annihilation and creation operators belonging to the same pair at equal times. Apparently, deviations from the correct, complete filtering scheme, cf. Fig. 2, are reasonably small, so that important effects of the filtering are correctly captured; except for the regime of very strong driving, where Rabi-like oscillations in the time-dependence gain strong influence on the measured $g^{(2)}(\tau = 0)$. The simple scheme suggests an intuitive understanding of the effect of filtering as a simple convolution of the unfiltered $G^{(2)}(\tau = t_2 - t_1)$ with the distribution function for the time difference $\tau$ due to the filtering effect on $t_{1/2}$. Note, that the time-difference distribution function is itself gained by convoluting the filter function in time domain with itself, so that $\tau$ is, in general, not distributed identical to the detection times $t_{1/2}$, but only for special cases of the filtering function.

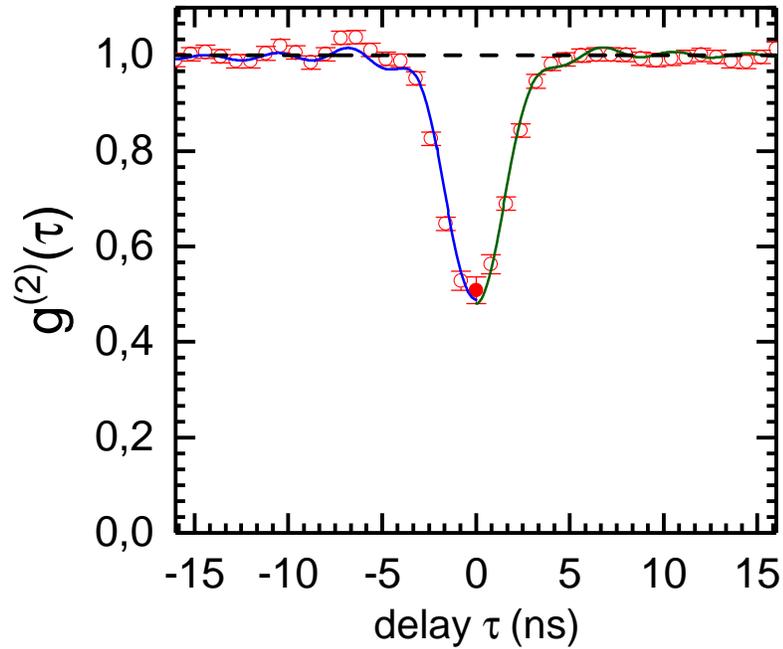

Figure 2: $g^{(2)}(\tau)$ at $n = 0.5$ photons. The green curve was computed using the 4-point filtering scheme, while the blue curve is the result of a simpler 2-point convolution. For this low photon number, the difference between the two curves is smaller than the error bars on the experimentally measured points (in red).

## V.  MEASUREMENT OF THE $g^{(2)}(\tau)$ FUNCTION

### A.  Principle of the measurement

Our measurement scheme is to process the small signals leaking out of the sample with standard microwave techniques (filtering, amplification and heterodyning), to digitize them with an acquisition card and to compute numerically the correlation functions relevant to characterize our single-photon source – the most important of them being the second-order coherence function:

$$g^{(2)}(t,\tau) = \frac{\langle \hat{a}^{\dagger}(t)\hat{a}^{\dagger}(t+\tau)\hat{a}(t+\tau)\hat{a}(t)\rangle}{\langle \hat{a}^{\dagger}(t)\hat{a}(t)\rangle\langle \hat{a}^{\dagger}(t+\tau)\hat{a}(t+\tau)\rangle}$$

Filtering and amplification are performed in multiple stages but can nonetheless be described as the action of a single effective amplifier of gain $G$, which adds a noise mode $\hat{h}$ in a thermal state at temperature $T_N$ to the input signal mode $\hat{a}$. The output of such an amplifier is then $\sqrt{G}\hat{a} + \sqrt{G-1}\hat{h}^{\dagger}$[12].

After digitization of this amplified signal we demodulate numerically its $I$-$Q$ quadratures. The $n$-th order moment of the complex envelope $S(t) = I(t) + jQ(t)$ is then a bilinear function of all the moments of $\hat{a}$ and of $\hat{h}$ up to order $n$[13]. By setting the bias voltage across the emitting squid to zero (the *off* position) and thus putting $\hat{a}$ in the vacuum state, we can measure independently the moments of $\hat{h}$. We then iteratively substract them from the moments of $S$ measured when the bias voltage is applied (*on* position) to reconstruct the moments of $\hat{a}$. Similarly, we can reconstruct $g^{(2)}(\tau)$ from the *on-off* measurements of all the correlation functions of $S(t)$ up to order 4.

In addition to this, by splitting the signal from the sample over two detection chains (channels "1" and "2") in a Hanbury Brown-Twiss setup, we can cross-correlate the outputs $S_1, S_2$ of the two channels to reduce the impact of the added noise on correlation functions. A model of this noise is thus needed to determine which combination of $S_1$ and $S_2$ is best suited to measure $g^{(2)}(\tau)$ accurately.

### B.  Model for the detection chain

The input-output formalism links the cavity operator $\hat{a}$ to the ingoing and outgoing transmission line operators $\hat{b}_{in}, \hat{b}_{out}$ by: $\sqrt{\kappa}\hat{a}(t) = \hat{b}_{in}(t) + \hat{b}_{out}(t)$, with $\kappa = 2\pi\nu_R/Q$ the energy leak rate (see Fig. 1). In our experimental setup, $\hat{b}_{in}$ describes the thermal radiation coming from the 50 $\Omega$ load on the isolator closest to the sample. This load being thermalized at 15 mK$\ll h\nu_R/k_B$, the modes impinging onto the resonator can be considered in their ground state and the contribution of $\hat{b}_{in}$ to all the correlation functions vanishes. We thus take $\hat{b}_{out}$ as being an exact image of $\hat{a}$, and all their normalized correlation functions as being equal.

As described before, the emitted signals are split between two detection chains, filtered, amplified, and mixed with a local oscillator before digitization. Each one of these steps adds a noise mode to the signal (FIG. 3). The beam-splitter right out of the resonator is implemented as a hybrid coupler with a cold 50 $\Omega$ load on its fourth port, which acts as an amplifier of gain $1/2$ and adds a noise mode in the vacuum state $\hat{h}_{bs}^{\dagger}$ to the signal[? ]. The different amplifying stages are summed up into one effective amplifier for each channel, with noise temperatures $T_N{}^1 = 13.5$ K and $T_N{}^2 = 14.1$ K respectively. The IQ mixer is used for heterodyning signals, i.e. shifting them to a lower frequency band where we can digitize them, and also adds at least the vacuum level of noise to the signals.

The last step, linear detection of the voltage $V_i(t)$ on channel $i$ by the acquisition card, is harder to model to the quantum level. After digitization, we process chunks of signal of length 1024 samples to compute the analytical signal $S_i(t) = V_i(t) + \mathfrak{H}(V_i)(t)$, with $\mathfrak{H}$ the discrete Hilbert transform. As computing the analytical signal from $V_i(t)$ accounts to measuring its two quadratures, which are non-commuting observables, quantum mechanics imposes again an added noise mode. We sum up this digitization noise with the heterodyning noise into a single demodulation noise $\hat{h}_{IQ_i}$ (Fig. 1).

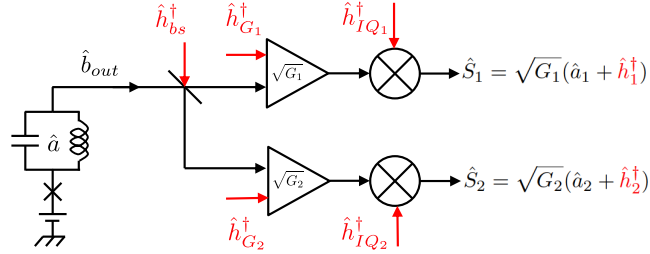

Figure 3: Detection chain model, taking into account all the added noise modes (in red).

In the end, we record measurements of $\hat{S}_1(t)$ and $\hat{S}_2(t)$, with $\hat{S}_i \propto \hat{a}_i + \hat{h}_i^\dagger$. Here $\hat{a}_i(t) \propto \hat{b}_{out}(t - \tau_i)$, where $\tau_i$ is the time delay on channel $i$. $\hat{h}_i$ is a thermal noise with an occupation number of about 65 photons, which summarizes the noises added by all the detection steps. In practice, the dominant noise contribution stems from the amplifiers closest to the sample. Note also that we do not consider here the effect of the finite bandpass of the filters, which complicates the link between $\hat{a}_i$ and $\hat{b}_{out}$.

### C. Computing correlations

From each chunk of signal recorded we compute a chunk of $S_i(t)$ of the same length 1024. We then compute the correlation functions we need as:

$$C_{X,Y}(\tau) = \langle X^*(t)Y(t+\tau) \rangle = \mathfrak{F}^{-1}(\mathfrak{F}(X)^* \mathfrak{F}(Y))$$

where $\langle ... \rangle$ stands for the average over the length of the chunk and $\mathfrak{F}$ is the discrete Fourier transform. Finally, we average the correlation functions from all the chunks and stock this result for further post-processing.

To illustrate how we reconstruct the information on $\hat{a}$ from $S_1, S_2$, let's consider the first order coherence function $g^{(1)}(\tau) = \frac{\langle \hat{a}^\dagger(t)\hat{a}(t+\tau) \rangle}{\langle \hat{a}^\dagger \hat{a} \rangle}$. We start with the product:

$$S(t)^* S(t+\tau) \propto \hat{a}^\dagger(t)\hat{a}(t+\tau) + \hat{h}(t)\hat{h}^\dagger(t+\tau) + \hat{a}^\dagger(t)\hat{h}^\dagger(t+\tau) + \hat{h}(t)\hat{a}(t+\tau)$$

We then make the hypothesis that $\hat{a}$ and $\hat{h}$ are independent and hence uncorrelated, which should obviulsy be the case as the noise in the amplifier cannot be affected by the state of the resonator. Then when averaging:

$$\langle \hat{a}(\tau)\hat{h}(t+\tau) \rangle = \langle \hat{a}(\tau) \rangle \langle \hat{h}(t+\tau) \rangle = 0$$

as there is no phase coherence in the thermal noise, i.e. $\langle \hat{h} \rangle = 0$. We then have:

$$\langle S(t)^* S(t+\tau) \rangle \propto \langle \hat{a}^\dagger(t)\hat{a}(t+\tau) \rangle + \langle \hat{h}(t)\hat{h}^\dagger(t+\tau) \rangle$$

Hence in the *off* position:

$$\langle S(t)^* S(t+\tau)\rangle_{off} \propto \langle \hat{h}(t)\hat{h}^\dagger(t+\tau)\rangle$$

and in the *on* position:

$$\langle S(t)^* S(t+\tau)\rangle_{on} \propto \langle \hat{a}^\dagger(t)\hat{a}(t+\tau)\rangle + \langle S(t)^* S(t+\tau)\rangle_{off}$$

such that:

$$g^{(1)}(\tau) = \frac{\langle S(t)^* S(t+\tau)\rangle_{on} - \langle S(t)^* S(t+\tau)\rangle_{off}}{\langle S^* S\rangle_{on} - \langle S^* S\rangle_{off}}$$

Now as we are considering states of the resonator with at most 1 photon, we typically have:

$\langle S^* S\rangle_{off} \simeq \langle S^* S\rangle_{on} \gg \langle S^* S\rangle_{on} - \langle S^* S\rangle_{off}$

Then any small fluctuation of the gain of the detection chain or of the noise temperature during the experiment reduces greatly the contrast on $g^{(1)}(\tau)$. We hence rely on the cross-correlation $X(\tau) = \langle S_1^*(t)S_2(t+\tau)\rangle$. Due to a small cross-talk between the two channels this cross-correlation averages to a finite value even in the *off* position, but which is 60 dB lower than the autocorrelation of each channel. We hence use:

$$g^{(1)}(\tau) = \frac{X(\tau)_{on} - X(\tau)_{off}}{X(0)_{on} - X(0)_{off}}$$

The same treatment allows to compute $g^{(2)}(\tau)$ with slightly more complex calculations. The classical Hanburry Brown-Twiss experiment correlates the signal power over the two channels, i.e. extracts $g^{(2)}(\tau)$ from $\langle S_1^* S_1(t) S_2^* S_2(t+\tau)\rangle$. The *off* value of this correlator is once again much bigger than the relevant information of the *on-off* part, and any drift of the amplifiers blurs the averaged value of $g^{(2)}(\tau)$.

To circumvent this difficulty, we instead use $C(t) = S_1^*(t)S_2(t)$ as a measure of the instantaneous power emitted by the sample, provided that the time delay between the two detection lines is calibrated and compensated for. We then have:

$$g^{(2)}(\tau) = \frac{\langle C(t)C(t+\tau)\rangle_{on} - \langle C(t)C(t+\tau)\rangle_{off}}{(\langle C\rangle_{on} - \langle C\rangle_{off})^2} - 2\frac{\langle C\rangle_{off}}{\langle C\rangle_{on} - \langle C\rangle_{off}}$$
$$- \frac{(X(\tau)_{on} - X(\tau)_{off})X(-\tau)_{off}}{(\langle C\rangle_{on} - \langle C\rangle_{off})^2} - \frac{(X(-\tau)_{on} - X(-\tau)_{off})X(\tau)_{off}}{(\langle C\rangle_{on} - \langle C\rangle_{off})^2} \quad (12)$$

## VI.  ELECTROMAGNETIC SIMULATIONS

The experiment can be schematically represented by figure 4, where the high impedance microwave mode we will use is represented in the green box as a $LC$ resonant circuit. Its resonant pulsation $\omega_0$ and characteristic impedance $Z_C$ are given by

$$Z_C = \sqrt{\frac{L}{C}} \quad ; \quad \omega_0 = \frac{1}{\sqrt{LC}}.$$

Aiming at a coupling strength $r \simeq 1$ at a frequency around 5 GHz, one gets, $Z_C \sim 2k\Omega$, $L \sim 60\,\text{nH}$ and $C \sim 15\,\text{fF}$.

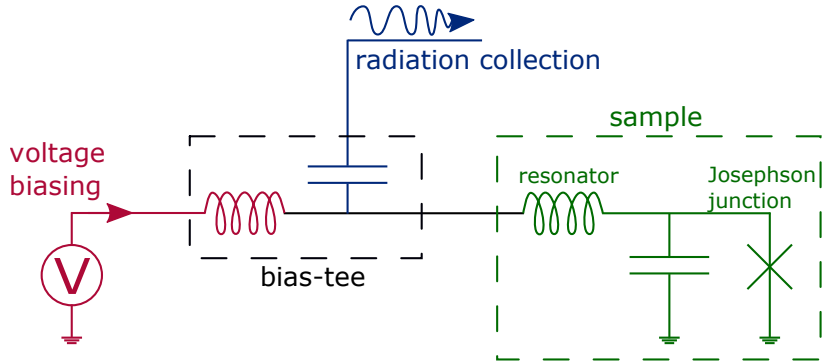

Figure 4: Schematic diagram of the experiment. The sample, represented by the green box consists in a Josephson junction galvanically coupled to a high impedance resonator consisting in a on-chip spiral inductor. It is connected to a DC biasing circuit, represented in red, and to a 50 Ω detection line, represented in blue, through a bias Tee.

In order to reduce the capacitance we fabricated the resonator on a quartz wafer, with a small effective permittivity $\varepsilon_r = 4.2$ (in comparison with 11.8 for silicon).

Planar coils [14, 15] offer an increased inductance compared to transmission lines resonators [16], and better linearity than Josephson based resonators [17]. They have, however, one main disadvantage : their center has to be connected either to the Josephson junction or to the detection line, using either bonding wires [15] or bridges [14]. Both solutions have a non negligible influence on the resonator. Bonding wires require bonding pad with typical size 50 $\mu$m, which increases the capacitance to ground, where as a bridge forms a capacitor with every turn of the coil that must be taken into account in the microwave simulations.

We chose to use an Aluminnum bridge, supported by a $> 1\mu$m BCB layer. BCB is a low loss dielectric which has been developed for such applications by the microwave industries which also has a relatively low permittivity.

A last parameter of the resonator that can be tuned is its quality factor Q. We consider a simple parallel LC oscillator with

$$Q = \frac{f_0}{\Delta f} = \frac{Z_C}{Z_{\text{Det}}},$$

where $Z_{\text{Det}}$ is the impedance of the detection line as seen from the resonator and $\Delta f$ the resonance bandwidth at -3dB. To tune $Q$, we can insert an impedance transformer between the 50Ω measurement line and the resonator and thus increase the effective input impedance, to decrease the quality factor.

In order to simulate our resonators, we use a high frequency electromagnetic software tool for planar circuits analysis : Sonnet. The system simulated by this software consists in several metallic layers separated by dielectrics as shown in Fig. 5.

Each metallic sheet layer contains a metallic pattern for the circuit, with strip-lines or resonators and can be connected to the other layers through vias. Dielectric layers properties and thickness can also be chosen.

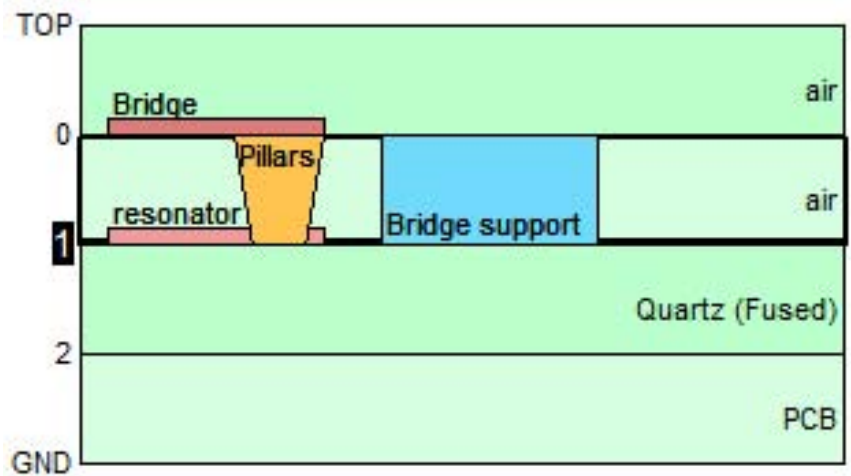

Figure 5: Sonnet schematic of the dielectric stack.

This stack is enclosed in a box with perfect metallic walls. The simulated device sees the outer world through ports that sit at the surface of the box, or are added inside the box, as probes shown in Fig. 6. Sonnet also allows us to insert lumped electric component in the circuit, between two points of the pattern.

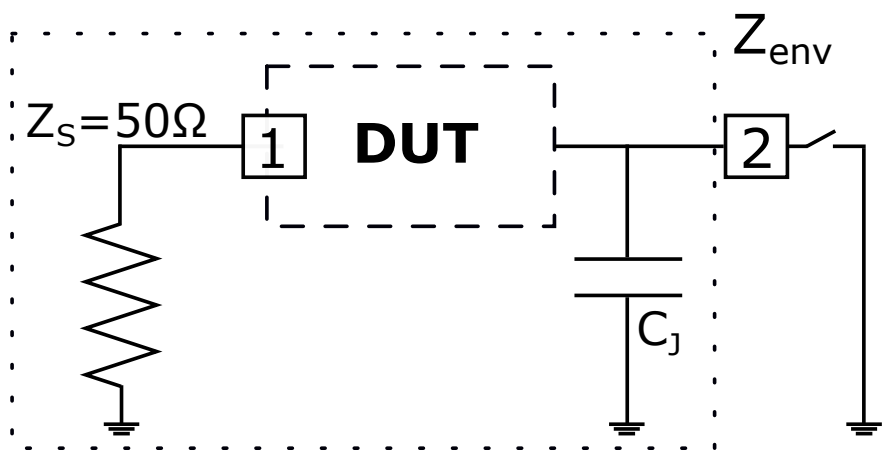

Figure 6: Sonnet port configuration.

As we are interested in the behavior of the environment seen by the junction, we will replace it by a port, which will act like a probe. We assume that the Josephson energy is small enough

for the admittance of the junction associated to the flow of Cooper pairs to be negligible ; we can thus model the junction as an open port. Furthermore, in order to take into account the junction's geometric capacitance, we add a discrete capacitor in parallel to ground as presented in Fig. 6. The other port of the resonator is model by a 50 $\Omega$ resistor, modeling the detection line.

Using the microwave simulation results, we predict the resonant frequency $f_0$, the impedance seen by the junction $Z_{\text{out 2}}$, the quality factor $Q$ and the environment characteristic impedance $Z_C$.

| Coil | nb of turns | line width | line space | bridge |
|------|-------------|------------|------------|--------|
| | 23.5 | 1 $\mu$ m | 2$\mu$m | BCB / 1.2$\mu$m |
| Results | $f_0$ | $\Delta f$ | $Z_C$ | Re($Z_{\text{env}}$)$_{\text{MAX}}$ |
| ($C_J = 2\,\text{fF}$) | 5,1 GHz | 60 MHz | 2,05 k$\Omega$ | 188k$\Omega$ |

Table I: Geometric parameters of the resonator and associated characteristics.

The corresponding schematic and result of simulations are shown in Fig. 7. Now, the lumped capacitor $C_J$ represents the capacitance of the Josephson junction alone, the rest of the capacitance being implemented by the surrounding ground.

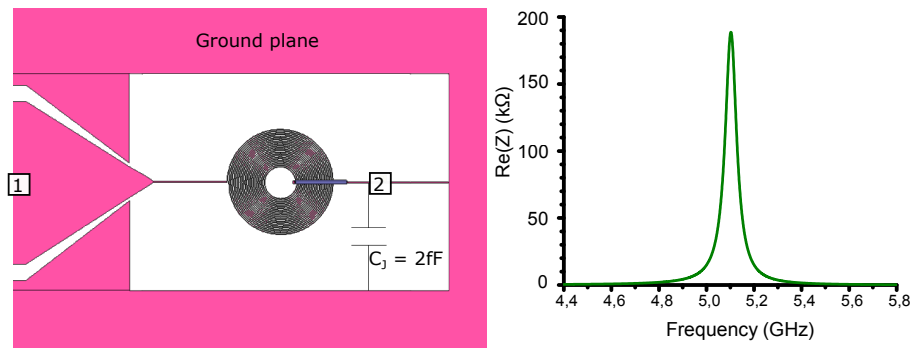

Figure 7: Final design drawing (left) and associated simulation result(right).

   *a. Computing current densities*   In order to understand the full resonator behavior, we have simulated current and charge densities at resonance, as shown in Fig. 8.

## A. Junction's capacitance influence

The Josephson junction's geometric capacitance $C_J$ is of the order of few fF (70 fF $.\mu m^{-2}$) [18] and is also part of the environment seen by the pure Josephson element according to

$$Z_{\text{env}}(\omega) = \frac{Z_{\text{circuit}}(\omega)}{1 + jC_J\omega Z_{\text{circuit}}(\omega)},$$

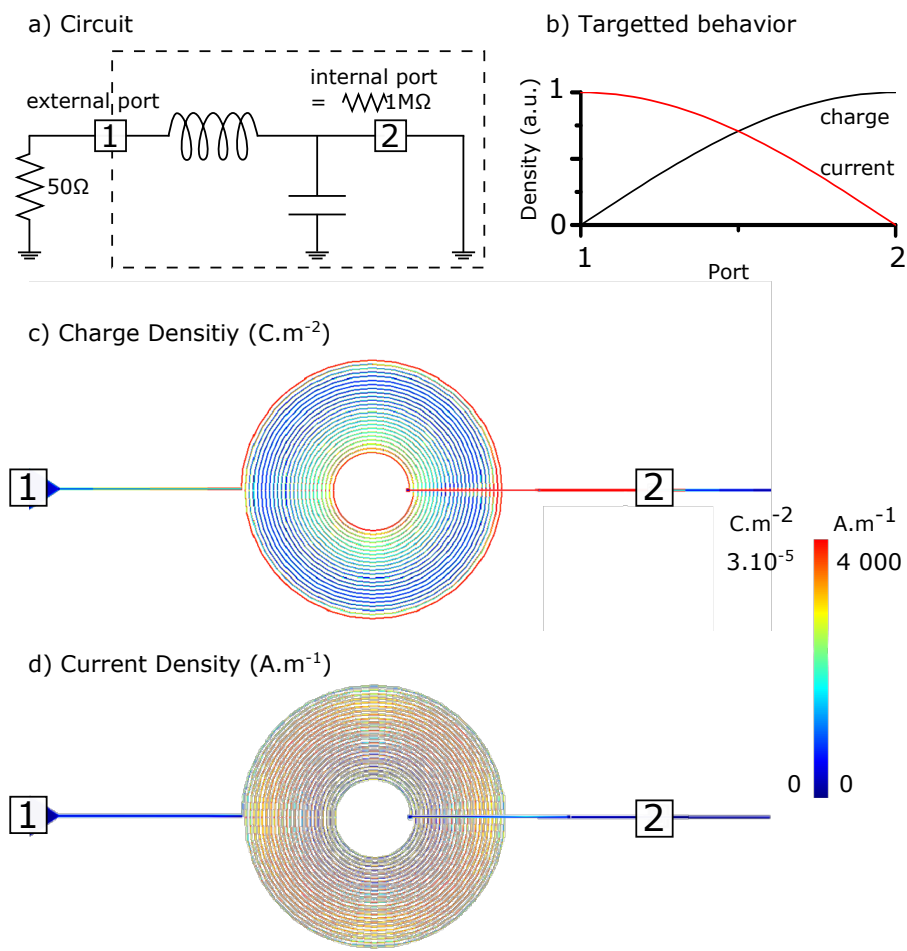

Figure 8: a) Our circuit has two ports. One external port "1" to be connected to the measurement chain can be modeled as a 50Ω load. The second port "2" is internal to the circuit and parametrized to mimic the Josephson junction open between the resonator and the ground plane probes the impedance seen by the future junction. These boundary conditions make us expect the resonator to behave like a λ/4 resonator. b) In a typical λ/4, the low impedance port 1 corresponds to a node in charge and an anti-node in current, while on the "open" side port 2, there is an accumulation of charges and no current. c) As expected, there is a charge accumulation on the high impedance side of the resonator. As the coil is used as an inductance but is also the capacitance of the circuit, there is an accumulation of charge at the periphery, i.e. in the first turn of the coil. d) There is indeed no current flowing through the high impedance side and we see an increase toward the low impedance port. in c) and d) the ground plane shown in fig. 7 is not represented here as it does not present peculiar current/charge density.

where $Z_{\text{circuit}}(\omega)$ is the impedance of the resonant circuit connected to the measurement line without junction.

By adding a discrete capacitor to ground $C = C_J$ at the junction's position in simulations and tuning its value, one can then observe in Fig. 9 that it is not negligible and must be taken into account.

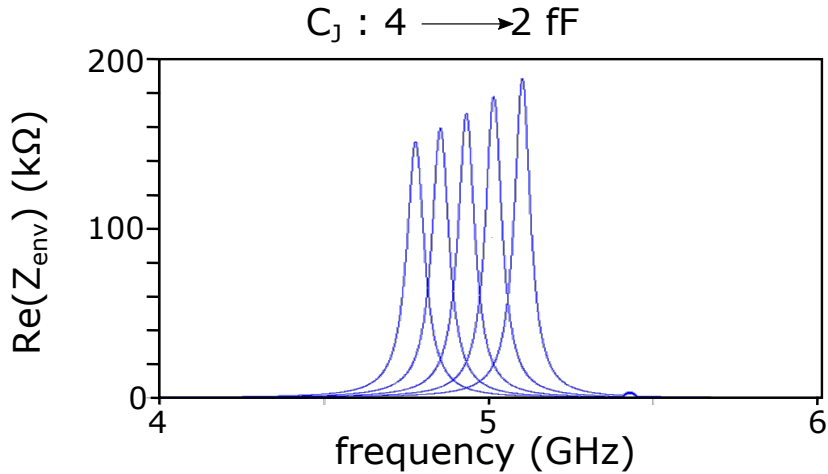

Figure 9: Real part of the impedance seen by the junction $\text{Re}[Z_{\text{env}}(\omega)]$ for different junction capacitances

As we aim at building a resonator with a capacitance around $15\,\text{fF}$, $C_J$ will account for 10 to 20% of the total capacitance of the circuit. As a consequence, both characteristic impedance and resonant frequency will be decreased by 5 to 10%.

### B. Tuning the bandwidth using quarter wavelength resonator

According to table I and Fig. 7, our resonator is expected to have a bandwidth of $\Delta f \sim 60\,\text{MHz}$, which is not much larger than the 3 MHz FHWM of the Josephson radiation due to low frequency voltage polarisation noise. It is thus useful to broaden this resonance while preserving the characteristic impedance and resonant frequency.

Keeping the same resonator geometry, one can enlarge its bandwidth by inserting a second resonator between it and the source to play the role of an impedance transformer (quarter wavelength). Doing so, we can increase the input impedance seen by the coil and broaden the resonance.

We have built this second stage of impedance transformer "on chip" between the measurement line (modeled by $Z_0$) and the coil, using a lossless coplanar waveguide (CPW) of length $l = \lambda/4$ according to :

This transformer is characterized by

$$Z_{\text{Det}} = \frac{Z_C^2}{Z_0},$$

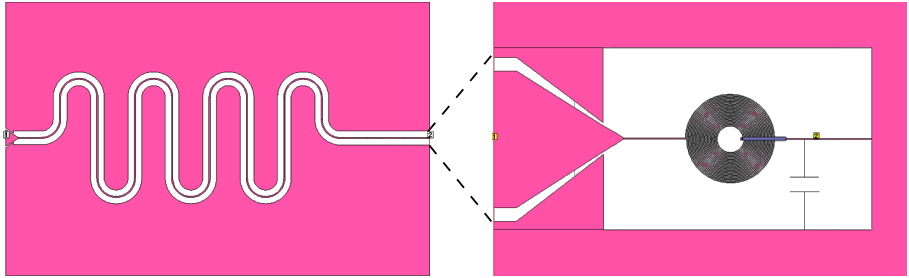

Figure 10: Circuit with an additional impedance transformer.

| $\Delta f$ (MHz) | $Z_{\text{Det}}$ | $Z_c, \lambda/4$ | width($\mu m$) | Gap ($\mu m$) |
|---|---|---|---|---|
| 60 | 50 | - | - | - |
| 100 | 100 | 70 | 25 | 10 |
| 300 | 400 | 140 | 10 | 50 |
| 500 | 600 | 173 | 5 | 67 |

Table II: Influence of an additionnal impedance transformer on the resonator bandwidth.

where $Z_C$ is the characteristic impedance of the line, $Z_{\text{Det}}$ the transformed detection impedance of the resonator and $Z_0$ the $50\Omega$ characteristic impedance of the detection line.

One can then choose the impedance seen by the coil (the impedance $Z_0'$ of the transformer) by tuning the characteristic impedance $Z_C$. To do so, textbook calculations allow to choose the good ratio between the width of the central conductor and the distance to ground plane on a particular substrate [19]. The bandwidth of the resonator, corresponding to an input impedance of $50\Omega$, is $60\,\text{MHz}$. By adding quarter wavelength transformers, we increase $\Delta f$ as listed in table VIB :

Using the quarter wavelength transformer simulations as a first block and the previous coil results as a second one, the full circuit was simulated, using the Sonnet "netlist" feature. Such a combination of previous simulations assumes no geometric "crosstalk" between the two resonators, which makes sense given that they are shielded from each other by ground planes. We obtained the results of Fig. 11.

We were then able to check that the $\lambda/4$ resonator has no influence on the characteristic impedance by extracting it for each design. In order to have different bandwidth, these four configurations were fabricated. In practice, the sample used in the experiments reported in the main text had a $70\,\Omega$ quarter wavelength inserted between the $50\,\Omega$ detection line and the planar inductor resonator.

## VII. FABRICATION

As mentioned above, we built the circuit of Fig. 12 on a $3 \times 10\,\text{mm}^2$ low permittivity quartz chip with a single input/output port adapted to a $50\Omega$ measurement line.

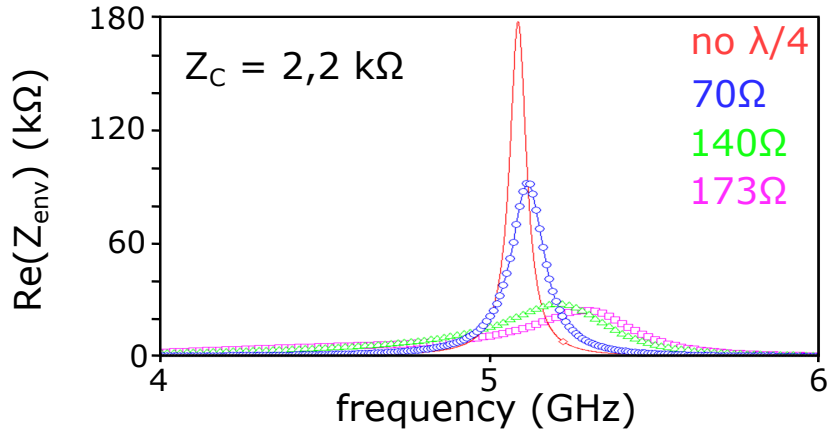

Figure 11: : initial simulation result and Netlist simulation results for the 3 λ/4 transformers of table VI B

Our fabrication process consists in 3 mains steps. First, we fabricate Niobium based coil and quarterwave impedance transfromers. Then, we connect the center of the coil to its periphery with a bridge and finally, as it is the most fragile element, we fabricate the Josephson junctions.

### A. Resonator: coil and λ/4

In order to be able to test the samples at 4K, we chose to built niobium based resonator. As Niobium is of a much better quality when sputtered than evaporated, we used a top down approach for this step.

A 100 nm a layer of Niobium was first deposited on a $430 \mu m$ thick Quartz wafer at 2nm/s using a dc-magnetron sputtering machine and then patterned by optical lithography and reactive ion etching (RIE).

In order to pattern the resonators, we used an optical lithography process. The classical optical lithography process used a resist thick enough so that all the niobium between the lines can be removed before all the resist is etched, the S1813 from Shipley.

It was spinned according to the following recipe:

1. 110°C prebake of the substrate on hot plate

2. Resist spinning : S1813, 4000 rpm 45" / 8000 rpm 15"

3. 2 min rebake on hot plate

Using these parameters, and performing interferometric measurements, we measured a resist layer of 1450 nm. The sample was then exposed with a Karl-Süss MJB4 optical aligner, with a dose of 150 mJ/cm$^{-2}$ (15 secs). Finally it was developed using microposit MF319 during 90 seconds and rinsed in deionized water for at least 1 min.

The next step of the process is the reactive ion etching of the niobium film : we used a mixture of CF$_4$ and Ar (20/10 cc) at a pressure of $50\mu$bar (plasma off) and a power of 50 W (209V) for 4

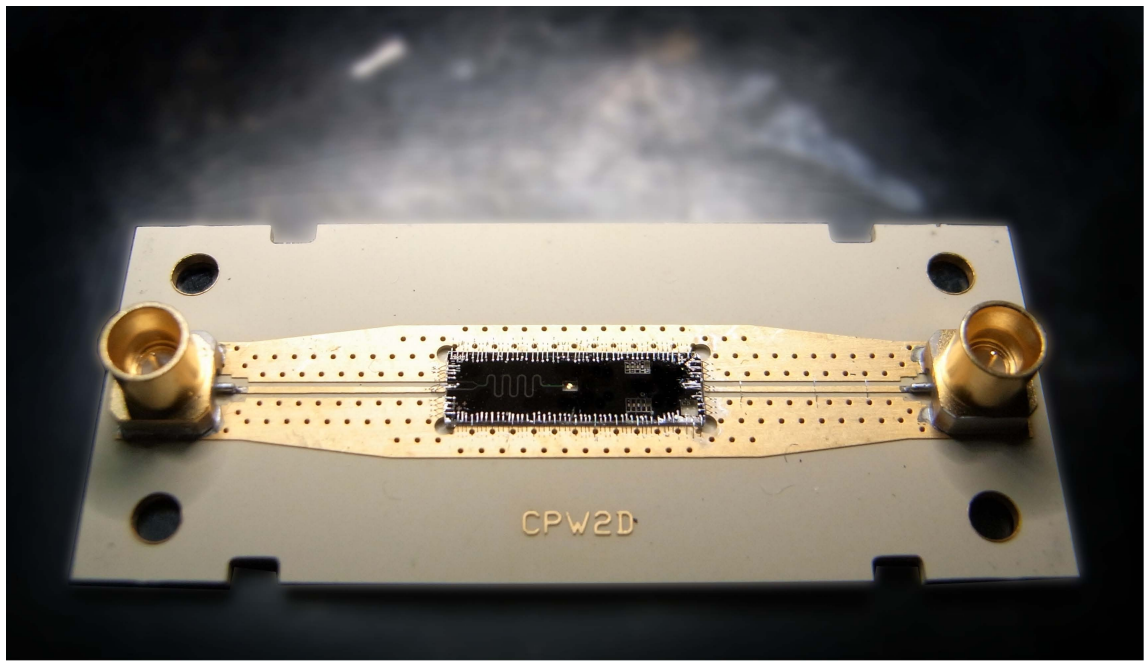

Figure 12: Photograph of the chip used for the experiments described in the main text.

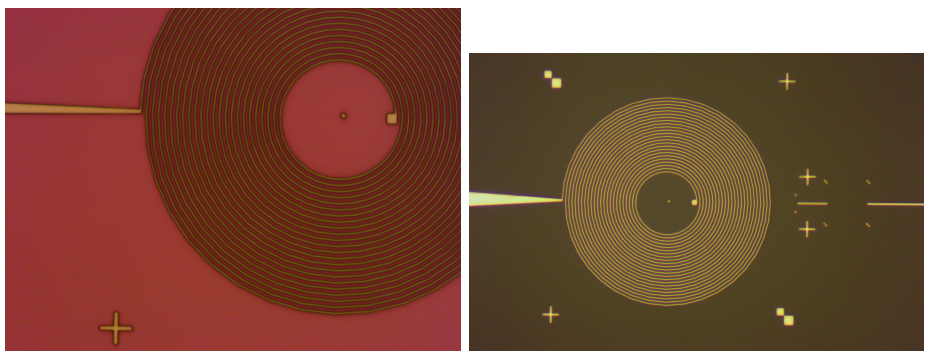

Figure 13: Fabrication of the coil. Left : photograph after optical lithography. Right : photograph after niobium etching.

minutes 45 seconds (150 nm). After this process, the sample was cleaned in 40°C acetone for 10 minutes to remove any resist residues and rinsed in IPA.

The quarter wave resonator was fabricated at the same time as the coil by Niobium etching.

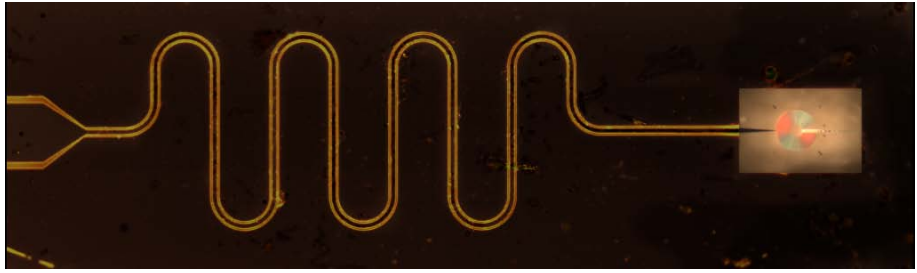

Figure 14: 70 Ω quarter wavelength impedance transformer and coil. The whole chip is $3 \times 10 \, \mathrm{mm}^2$.

**B.   Bridge**

As we decided to use a dielectric spacer to support the bridge, we added two additional steps to the fabrication process. One for the dielectric spacer, the second one for the brdige itself. One of the main difficulties of these steps is that, as the pads to connect the bridge is small, they require very precise alignment.

*a.   Dielectric support*   We chose to work with polymers derived from B-staged bisbenzocyclobutene, sold as Cyclotene 4000 by Dow Chemicals and choose the lower viscosity, in order to obtain a spacer between 0.8 and $1.8 \mu m$ thick: XU35133. The process was performed according to the following recipe:

1. 2 minutes prebake at 110˚ C

2. Primer AP 3000 rpm 30 secs

3. BCB XU : 3000rpm, 45secs/ 8000 rpm 15 secs

4. 3 minutes rebake @80˚ C

Using this technique, we obtained 1650 nm thick layers. The sample was then exposed with the MJB4 optical aligner, with during 3 seconds. The development of this resist is quite difficult as it is not dissolved by acetone:

1. 30 secs on hot plate (70˚ C) : to avoid that the bridge flows

2. DS 3000 rinsing for 1 minute

3. TS 1100 rinsing for 30 seconds

4. 1 min rinsing in deionized water

5. the sample was then dried while spinning

In order to obtain a flat surface and remove all resist residues, an RIE $SF_6$ /$O_2$ etching was performed for 30 seconds (20/2 cc, $10\mu$bar, 50W) as shown in Fig 15. Finally, the sample was rebaked during 30 minutes at 190˚C to stabilize the resist.

Stepper measurement of BCB resist profile before and after etching

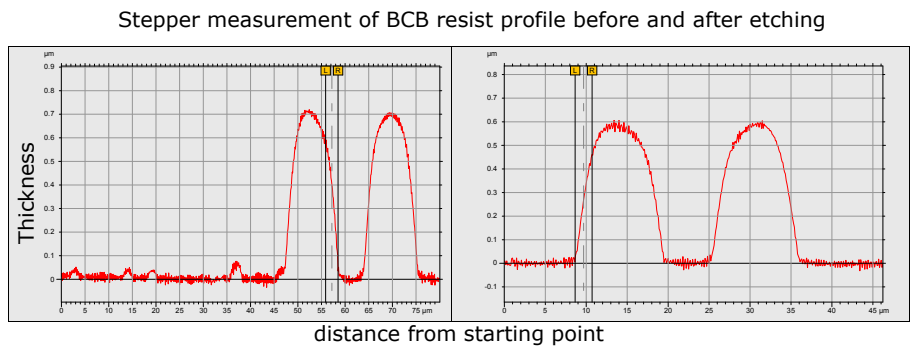

Figure 15: The flatness of the sample was measured with a stepper. After 30 seconds of $SF_6$/$O_2$ etching, resist residues have disappeared

*b. Bridge's line*   As the spacer is quite thick, this step requires a thicker resist. We used AZ5214 and obtained a $1.5\mu m$ layer of resist according to the recipe:

1. 72˚C prebake on hot plate

2. microposit primer : 6000rpm for 30 seconds

3. AZ5214 : 4000 rpm during 60s, 8000rpm during 10s

4. 2min rebake at 100˚C with a bescher on top of the sample

The sample was then aligned and exposed during 7s using the MJB4. As the AZ5214 is a negative resist which can be reversed, we rebaked the sample for 3min at 120˚C and performed a flood exposure for 25 seconds. The development was then performed using diluted AZ 400K with deionized water (1:4) for 1 min. Finally, the BCB was covered with a 200 nm layer of aluminum after 12 seconds Argon etching to ensure good contacts with the coil.

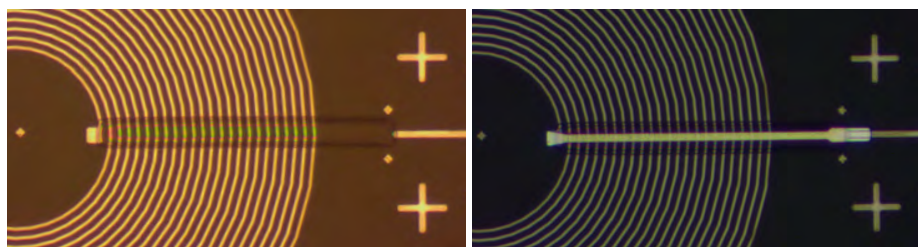

Figure 16: Left: in a first step, a BCB brick is deposited with optical lithography. Right: in a second step the core of the coil is connected with an aluminum bridge.

## C. Josephson junction

As explained in the main text, for $r \simeq 1$, strong anti-bunching effects are expected when the resonator is, in average, almost empty. The maximum photon emission rate is given by

$$\dot{n} = \frac{\text{Re}[Z(\omega_0)]I_0^2}{2\hbar\omega_0},$$

from which the mean occupation number $n$ can be deduced by:

$$\dot{n} = \frac{n}{\Gamma},$$

with $\Gamma = 2\pi\,\text{HMBW}$, the leaking rate of the resonator. In order to estimate the targeted resistance of the junction, one uses the Ambegaokar-Baratoff formula and Josephson relations (taking into account that DCB will renormalizes $E_J$ by a factor of $\exp(-\pi Z_c/2R_Q)$):

$$I_0 = \frac{\pi\Delta}{2\,\text{e}R_N}, \qquad\qquad E_J = \frac{\varphi_0 I_C}{2\pi}.$$

In order to be able to tune $E_J$ with a magnetic field, a SQUID geometry is used for the josephson junction: two junctions are placed in parallel to form a loop, which behaves as a single effective junction tunable with the external magnetic flux applied to the loop.

As a small capacitance is required for the resonator, junctions must be as small as possible, but big enough to be reproducible and lead to a good symmetry between the two branches of the SQUID. Assuming a symmetry of 90%, $E_J$ can then be reduced by a factor of 10 tuning the flux with a little coil on top of the sample.

Assuming a bandwidth $\Delta\omega \sim 100\,\text{MHz}$, a characteristic impedance $Z_C \sim 2k\Omega$, a critical current $I_0$ of $1\,\text{nA}$ and a symmetry of 90%, one can estimate the minimal amount of photon in resonator :

$$n = 1/100.\frac{\dot{n}}{\Gamma} = \frac{Z_C I_0^2}{2h(\Delta\omega)^2.100}e^{-\pi Z_c/R_Q} \sim 0.5$$

with $\Delta\omega$ the half maximum bandwidth of the resonator (FWHM). These parameters require a normal state resistance for the SQUID of $R_N \sim 300k\Omega$.

*a. Fabrication principle* Samples are made of aluminum based tunnel junctions, fabricated by double angle evaporation through a suspended shadow mask, using the standard Dolan technique [20]. By adjusting the angles of evaporation, two adjacent openings in the mask can be projected onto the same spot, creating an overlay of metallic films as shown in fig. 17. The first film is oxidized before the second evaporation to form the tunnel barrier.

In order to have reproducible as well as small junctions, we used a cross shape as shown in Fig. 18.

*b. SQUID fabrication* PMMA/PMGI resist bilayer spining :

1. 2 min rebake at 110° C

2. Ti prime 6000 rpm 30 secs

3. PMGI SF8 : 3000rpm, 45secs/ 6000 rpm 15 secs ($\approx 613 \pm 15$nm)

4. 5min rebake @170°C with bescher

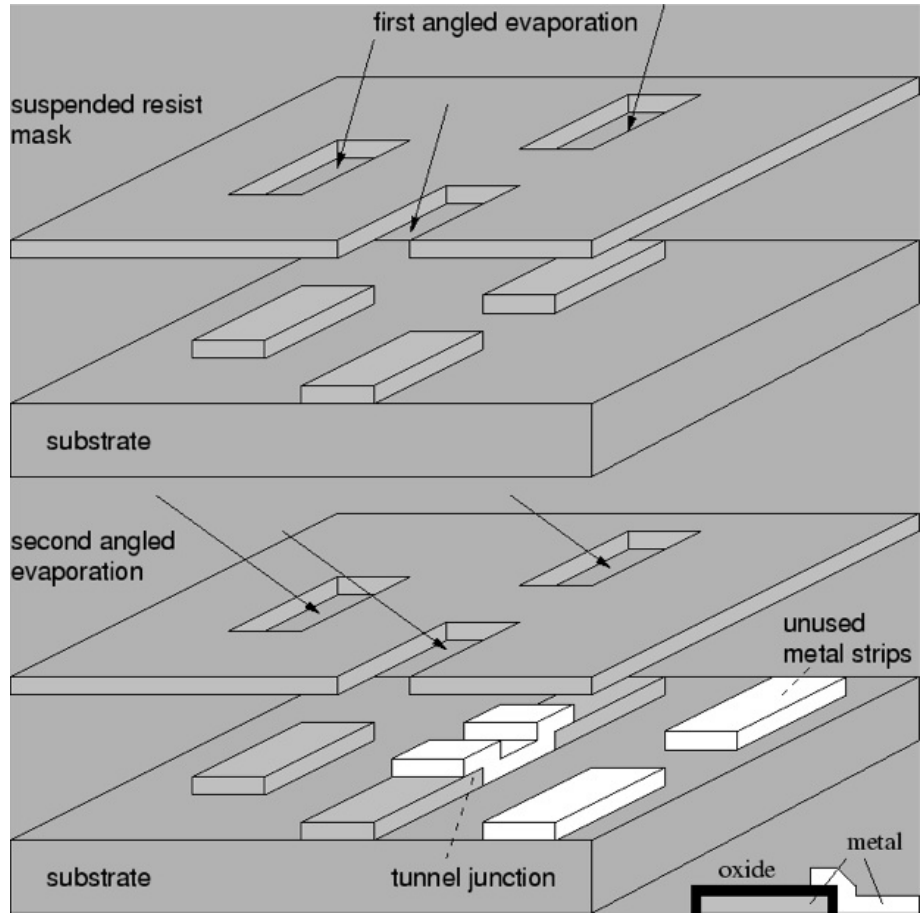

Figure 17: Double angle evaporation principle: two metallic layers are evaporated onto the same spot, creating an overlay of metallic films. As the first layer was oxydized, the two electrodes are separated by an insulator and form a Josephson SIS junction.

5. PMMA A6 : 6000rpm, 60secs ($\approx$ 253$\pm$21nm)

6. 15 min rebake @ 170°C (with bescher)

As the quartz is very sensitive to charging effects, we placed an additional 7nm layer of aluminum of top of the resist to evacuate charges during EBL. The full wafer was then covered by a thick layer of UVIII resist which can be removed in IPA and sent to IEF for dicing. Actually, as the Quartz substate has an hexagonal symmetry, it cannot be cleaved.

We then performed EBL on single chips using an FEI XL30 SEM with a dose of 300 $\mu$C.cm$^{-2}$ at 30 kV. The focus was tuned a three point on the sample using 20 nm gold colloids.

The development process then consisted in :

1. 35 secs MIF 726, 15 secs ODI to remove the aluminum layer

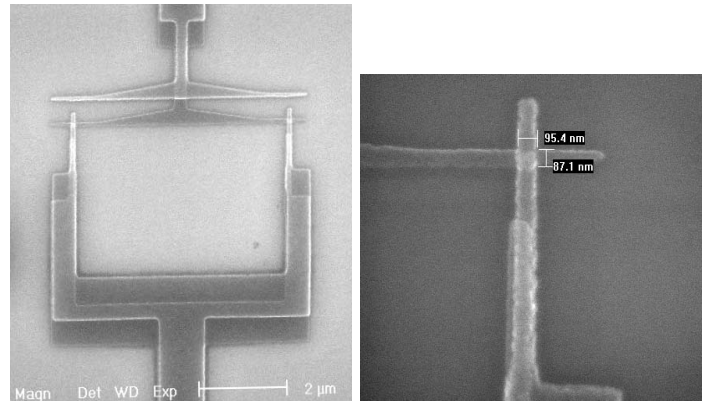

Figure 18: Left: SEM image of the SQUID. Right: zoom on one of the Josephson junction with size $95 \times 87 \, \text{nm}^2$

2. 60 secs MIBK + IPA (1:3), 30 secs IPA, 15 secs ODI to open the Josephson junction patterns

3. 25 secs MIF 726, 1min ODI, 15 secs ethanol to have a nice undercut

**Double oxidation junctions**  Finally, we deposited and oxidized aluminum to form highly resistive Josephson junctions using double angle evaporation technique. In order to fabricate very resistive Josephson junctions, the group of J.P. Pekola [18] raised the idea of oxidizing not one layer of aluminum but to do it twice. By evaporating an additional subnanometer thick layer of Al immediately after oxidizing the first layer, and oxidizing this fresh very thin layer, one thus obtain thicker barriers.

The key parameter of this recipe is the thickness of the intermediate thin Al layer. As it will be completely oxidized we can achieve resistances up to $1 M\Omega$ with 0,4nm. Using this process, the surfacic capacitance of the junction is estimated to $70 \, \text{fF} \, / \mu m^2$ i.e. $\sim 2 \, \text{fF}$ for the SQUID.

1. Argon ion milling 2x10 secs / 3 mA

2. -24˚ : 20 nm Al @ 1 nm.s-1

3. O2/Ar (15/85 %) oxidation 300 mbar during 20 min

4. 0.25 nm Al @ 0.1 nm.s-1

5. O2/Ar (15/85 %) oxidation 667 mbar during 10 min

6. 24˚ : 80 nm Al @ 1 nm.s-1

The lift-off of the resist was done by putting the sample in 60˚C remover-PG during 40 minutes. In order to get uniform resistance values and limit Josephson junctions aging, they were rebaked on a hot plate at 110˚C during one minute.

The chip was then stuck on the PCB with UVIII resist and bonded to the single input/ output port using aluminum wires as shown in Fig. 12.

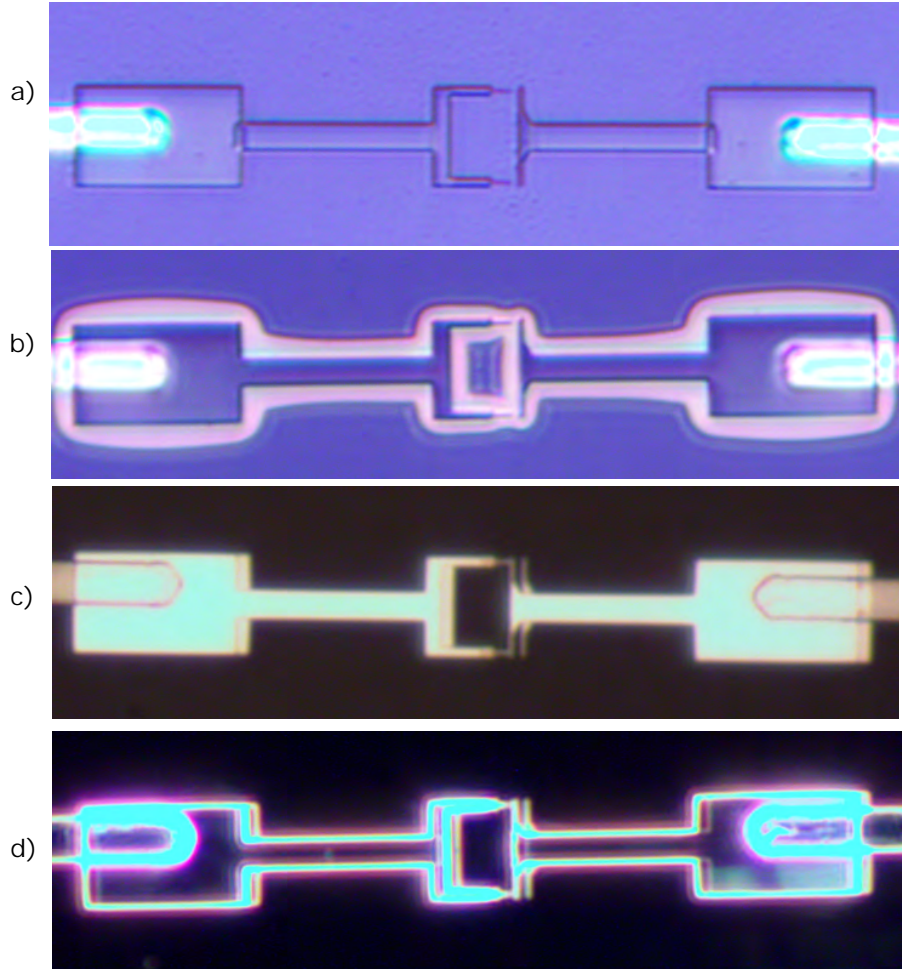

Figure 19: Josephson junction fabrication steps : a) Josephson junction shape : PMMA development b) Undercut : PMGI development c) & d) optical microscope view of the junctions after lift-off

## VIII. CALIBRATION PROCEDURE

To calibrate in-situ the gain $G$ of the detection chain and the impedance $Z(\nu)$ seen by the junction, we measure the power emitted by the junction in two different regimes. First, we bias the junction well above the gap voltage $V_{\mathrm{gap}} = 210 \ \mu V$ and measure the voltage derivative of the quasiparticle shot noise power spectral density $\mathcal{P}(\nu)$:

$$\frac{\partial \mathcal{P}(\nu)}{\partial V} = \frac{2 e R_t \mathrm{Re} Z(\nu)}{|R_t + Z(\nu)|^2}, \tag{13}$$




\begin{thebibliography}{42}
\bibitem[{not()}]{noteEJ}
\bibinfo{note}{The flux jumps due to vortex depinning were slow enough that we
  could compensate for them manually. We could thus obtain the total flux
  dependance of the emitted power, whose fit with a sinusoidal law yields the
  same value for $E_J$ within experimental errors and a negligible asymmetry.}
\end{thebibliography}
\end{document}